\documentclass[preprint,11pt]{elsarticle}

\usepackage{amssymb}
\usepackage{graphicx,subfigure}

\journal{Information Sciences}

\newtheorem{property}{Property}
\newtheorem{corollary}{Corollary}
\newtheorem{lemma}{Lemma}
\newtheorem{theorem}{Theorem}
\newtheorem{definition}{Definition}

\newtheorem{remark}{Remark}

\newtheorem{example}{Example}

\newcommand{\Sast}[0]{{\Sigma}^\ast}
\newcommand{\lder}[1]{\partial_{#1}}

\begin{document}

\hyphenation{non-mem-ber-ship}

\begin{frontmatter}



\title{Recognizability of languages via deterministic finite automata with values on a monoid:\\
General Myhill-Nerode Theorem}


\author{Jos\'e R. Gonz\'alez de Mend\'{i}vil}
\ead{mendivil@unavarra.es}
\author{Federico Fari\~na Figueredo}
\ead{fitxi@unavarra.es}

\address{Departamento de Estad\'istica, Inform\'atica y Matem\'aticas}
\address{Universidad P\'ublica de Navarra}
\address{31006 Pamplona (Spain)}

\begin{abstract}
This paper deals with the problem of \emph{recognizability} of functions $\ell:\Sast \rightarrow M$ that map words to values in the support set $M$ of a monoid $(M, \cdot, 1)$. These functions are called $M$-languages. $M$-languages are studied from the aspect of their recognition by deterministic finite automata whose components take values on $M$ ($M$-DFAs). The characterization of an $M$-language $\ell$ is based on providing a right congruence on $\Sigma^\ast$ that is defined through $\ell$ and a \emph{factorization} on the set of all $M$-languages, $L(\Sast,M)$ (in short $L$). A \emph{factorization} on $L$ is a pair of functions $(g,f)$ such that, for each $\ell\in L$, $g(\ell)\cdot f(\ell)=\ell$, where $g(\ell)\in M$ and $f(\ell)\in L$. In essence, a factorization is a form of common factor extraction. In this way, a general \emph{Myhill-Nerode theorem}, which is valid for any $L(\Sast, M)$, is provided. Basically, \emph{$\ell\in L$ is recognized by an $M$-DFA if and only if there exists a factorization on $L$, $(g,f)$, such that the right congruence on $\Sast$ induced by the factorization $(g,f)$ and $f(\ell)\in L$, has finite index}.
This paper shows that the existence of $M$-DFAs guarantees the existence of natural non-trivial factorizations on $L$ without taking account any additional property on the monoid. In addition, the composition of factorizations is also a new factorization, and the composition of natural factorizations preserves the recognition capability of each individual natural factorization.
\end{abstract}


\end{frontmatter}

\section{Introduction}\label{sec:Introduction}


In formal languages and automata \cite{Hopcroft2007}, the \emph{Myhill-Nerode Theorem} \cite{Myhill1957} \cite{Nerode1958} provides necessary and sufficient conditions for a language to be recognized by a deterministic finite automaton (DFA). In that case, the language is said to be \emph{regular}. Specifically, given a language $\ell\subseteq \Sast$ (where $\Sast$ denotes the set of all finite words on an alphabet $\Sigma$) a right congruence relation $\equiv_\ell$ on $\Sast$ is defined in terms of the language $\ell$, but with no regard to its representation. The recognizability of a language $\ell$ by a DFA is established by proving that: \emph{$\ell$ is a regular language if and only if $\equiv_\ell$ has finite index}. It is worth of mention that the states of the DFA based on $\equiv_\ell$, that recognizes $\ell$, are the equivalence classes of that congruence. Furthermore, this DFA is minimal what makes Myhill-Nerode theorem to be considered in the topic of automata minimization.

In order to cope with different domains of practical applications, researchers have presented in the literature effective generalizations of automata using a wide diversity of algebraic structures. \emph{Fuzzy automata} and \emph{weighted automata} are ones of the best-known studied generalizations of automata \cite{Mordeson2002}\cite{Droste2009}. For weighted automata, values on the transitions of those automata are usually taken from semirings \cite{Mohri1997}, hemirings \cite{Droste2013}, or strong bimonoids \cite{Ciric2010}\cite{Droste2010}. For fuzzy automata, values on transitions are taken from certain ordered structures like lattice-ordered monoids \cite{Li2005}, lattice-ordered structures \cite{Li2011}, complete distributive lattices \cite{Belohklavek2002}, general lattices \cite{Li2011}, or complete residuated lattices \cite{Ignjatovic2008}\cite{Qiu2001}\cite{Qiu2002}.


\emph{Weighted deterministic finite state automata} are called \emph{(sub)sequential transducers} \cite{Eilenberg1974}\cite{Sakarovitch2009}. These transducers recognize languages called \emph{(sub) sequential rational functions}. Characterization of the (sub)sequential rational functions in terms of a congruence relation in the flavour of Myhill-Nerode theorem
has been studied for different special cases of monoids like free monoids \cite{Sakarovitch2009}, $(\mathbb{R}_0^+, +, 0)$  \cite{Mohri2000}, \emph{gcd monoids} \cite{Souza2000}, or monoids based on \emph{sequentiable structures} \cite{Gerdjikov2017}.

It is remarkable that Gerdjikov has provided five algebraic axioms \cite{Gerdjikov2018}\cite{Gerdjikov2018b}, based on the relation \emph{divisor of}\footnote{Given a monoid $(M,\cdot, 1)$. For $a, b \in M$, $a$ is \emph{divisor of} $b$ if there is $c\in M$ such that $b=a\cdot c$.} on the support set $M$ of the monoid, for characterizing a wide class of monoids. Those axioms are satisfied by groups, free monoids, sequentiable structures, tropical monoids (including $(\mathbb{Q}_0^+,+,0)$), and gcd monoids \cite{Gerdjikov2018b}. However, the axioms \emph{left cancellation} and \emph{right cancellation} considered in \cite{Gerdjikov2018b} avoid the existence of a \emph{zero} element in $M$.

In fuzzy languages and automata \cite{Zadeh1965}, Ignjatovi\'c et al. \cite{Jelena2010} have proposed a Myhill-Nerode type theory for fuzzy languages with membership values in an arbitrary set with two distinguished elements $0$ and $1$, $(M,0,1)$, which are needed to take common (\emph{crisp}) languages in consideration. These fuzzy languages are studied in \cite{Jelena2010} from the aspect of their recognition by \emph{crisp determinist fuzzy automata}. A crisp deterministic fuzzy automaton is simply an ordinary deterministic automaton equipped with a fuzzy subset of final states \cite{Belohklavek2002}. It is worth of mention, that the Myhill-Nerode theorem for fuzzy languages provided by Ignjatovi\'c \cite{Jelena2010} includes the recognizability of fuzzy languages with membership values on some well-known structures: G\"odel structure \cite{Shen1996} \cite{Mordeson2002}(Chap.7) \cite{Petkovic2005}; distributive lattices \cite{Rahonis2009}; finite monoids \cite{Bozapalidis2008}; and general lattices \cite{Li2011}. However, all these previous works in fuzzy languages have a severe limitation: they only consider recognizability of fuzzy languages of \emph{finite rank}. In fact, \emph{a fuzzy language is recognized by a crisp deterministic fuzzy automaton if and only if it has a finite rank and all its kernel languages are recognizable} (see Theorem 4.3 in \cite{Jelena2010}).

In order to circumvent this restriction, a generalization of Myhill-Nerode theorem for fuzzy languages, having finite or infinite rank, is introduced in \cite{mendivil2016} for fuzzy languages based on continuous triangular norms (t-norms)\cite{Klement2000}, $([0,1],\otimes, 1)$. In that paper, the characterization of fuzzy languages is based on a right congruence defined by using the notion of \emph{factorization}, in particular \emph{maximal factorization}.


In essence, a factorization is a form of common factor extraction. Specifically, a \emph{factorization} is a pair of functions $(g,f)$ such that they satisfy $\ell= g(\ell)\otimes f(\ell)$ for any fuzzy language $\ell$. This notion was initially introduced in weighted automata by Kirsten and M\"{a}urer \cite{Kirsten2005} and applied to the area of fuzzy automata in order to develop efficient constructions for determinization and minimization of fuzzy automata. In fact, factorizations have allowed researchers to provide minimization algorithms \cite{Mohri1997}\cite{Eisner2003} and determinization methods for weighted automata \cite{Kirsten2005}. In the context of fuzzy automata, factorizations have been studied to obtain determinization methods \cite{Mendivil2014}\cite{Mendivil2014b}\cite{Stanimirovic2017}, and minimization algorithms for fuzzy automata \cite{mendivil2018}\cite{mendivil2018b}\cite{Stanimirovic2017b}.

Although \cite{mendivil2016} deals with characterizing fuzzy languages with infinite rank, that work is somewhat restrictive because it assumes \emph{zero-divisor-free} t-norm based monoids and maximal factorizations. In general, that property on the monoid is a necessary condition for the existence of maximal factorizations as it has been recently proved in \cite{Gerdjikov2019}.


The motivation to write this paper is to study the recognizability of functions $\ell:\Sigma^\ast \rightarrow M$ defined for any arbitrary monoid $(M, \cdot, 1)$. In other words, we do not consider any additional property on the monoid when we address the recognizability problem unlike other previous research works that have been referenced in this introductory section. In order to get a compact presentation of our results, we assume that each $\ell:\Sigma^\ast \rightarrow M$ is a total function. We call these functions $M$-languages. As in previous works, recognizability is based on deterministic finite automata, but in this case, their components take values on $M$ ($M$-DFAs). Let $L(\Sast, M)$ be the set of all $M$-languages. Our main objective is to provide a general Myhill-Nerode theorem valid for any $L(\Sast, M)$. We use factorizations on $L(\Sast, M)$ in order to get a characterization of $M$-languages. The characterization of an $M$-language $\ell\in L(\Sast, M)$ is based on a right congruence on $\Sast$ induced by a factorization on $L(\Sast, M)$, $(g,f)$, and $\ell$. That congruence is denoted by $\equiv^{(g,f)}_\ell$ in the paper.

The general Myhill-Nerode theorem presented in this paper establishes that \emph{$\ell\in L$ is recognized by an $M$-DFA if and only if there exists a factorization on $L$, $(g,f)$, such that the right congruence on $\Sast$ induced by the factorization $(g,f)$ and $f(\ell)\in L$, has finite index}.

The proof is made by construction. In this way, we prove that if $\equiv^{(g,f)}_{f(\ell)}$ has finite index then there exists an $M$-DFA that recognizes the $M$-language $f(\ell)$, and, as $(g,f)$ is a factorization, $g(\ell)\cdot f(\ell)=\ell$, which implies that $\ell$ is also recognized by such an $M$-DFA. In the other direction, if an $M$-language $\ell$ is recognized by some $M$-DFA, $A$, then there exists a \emph{natural factorization} on $L(\Sast,M)$ induced by $A$, $(g_A,f_A)$, such that $\equiv^{(g_A,f_A)}_{f_A(\ell)}$ has finite index. The construction of a factorization induced by an $M$-DFA is not trivial and requires the notion of \emph{transition-equalized} automata. However, it is not necessary to construct explicitly such kind of $M$-DFAs to get their factorizations \cite{fitxi2019}.

Each possible factorization on $L(\Sast,M)$ has its own \emph{recognition capability}. Thus, we study the recognition capability of
three particular cases: trivial factorization, maximal factorizations and natural factorizations. The formulation of factorizations on $L(\Sast,M)$ allows us to define the composition of factorizations to form new factorizations. In this way, we prove that the recognition capability of the composition of natural factorizations preserves the recognition capability of each individual natural factorization.

The rest of the paper is organized as follows. Section \ref{sec: preliminaries} and section \ref{sec: factorizations} present a formal framework to define the operations and main properties of factorizations. Section \ref{sec: M-dfa} is a short introduction to $M$-DFAs. In section \ref{sec: recognizability}, we introduce the sufficient condition for recognizability of $M$-languages based on factorizations and the definition of right congruence based on factorizations. This section presents the main properties of the $M$-DFA based on this kind of congruences. Section \ref{sec: natural factorizations} is devoted to natural factorizations, i.e., the factorizations induced via $M$-DFAs. The properties of the $M$-DFA constructed under a natural factorization are also included in this section. The general Myhill-Nerode theorem is proved in section \ref{sec: main result}.
The recognition capability of a factorization is defined in section \ref{sec: recogn capability} and three cases of study are considered: trivial factorization, maximal factorizations and composition of natural factorizations. Finally, some concluding remarks end the paper.

\section{Preliminaries}\label{sec: preliminaries}

Let $f_1:A\rightarrow B$ and $f_2:B \rightarrow C$ be two well defined functions. In this paper, the composition of the functions $f_1$ and $f_2$ is denoted by $f_2 \circ f_1$ and defined by $(f_2 \circ f_1)(a)= f_2(f_1(a))$ for any $a\in A$. We will assume that every function is a total function.

Let $(M,\cdot, 1)$ be an arbitrary monoid where $M$, $\cdot$ and $1$ represent the support set, the multiplication operation, and the identity element of the monoid respectively. In general, we identify each monoid with the name of its support set.

Let $\Sigma$ be a finite alphabet of symbols. The set $\Sast$ denotes the set of all finite words over $\Sigma$. We use $\varepsilon$ to represent the empty word. Then, $\Sast$ is the free monoid generated by $\Sigma$ under the operation of concatenation. Let us consider functions from $\Sast$ to $M$, i.e., $\ell: \Sast \rightarrow M$. The set $L(\Sast, M)=\{\ell~|~\ell: \Sast \rightarrow M\}$ is the set containing all those kind of functions. We do not write $(\Sast, M)$ when it is clear in the context of discussion. Thus, $L(\Sast, M)$ is simply denoted by $L$. We consider that $L$ represents the set of all possible $M$-languages on $\Sigma$. In fact, if the monoid $M$ is the elemental Boolean monoid then, any $\ell\in L$ may be interpreted as the characteristic function of an ordinary language, i.e., a subset of $\Sast$.

In the following, we extend the multiplication operation $\cdot$ of the monoid $M$ to different contexts. The context for the symbol $\cdot$ will clarify its interpretation. Given $\ell \in L$ and $m \in M$, $m\cdot \ell\in L$ is defined as $(m\cdot \ell)(\gamma)= m\cdot \ell(\gamma)$ for every $\gamma \in \Sast$. Obviously, $(n\cdot m)\cdot \ell= n\cdot (m\cdot \ell)$ for any $n,m \in M$. For any function $r:\Sast \rightarrow \Sast$ (a word-transformation), any $\ell \in L$, and any $m\in M$, the next property holds:
\begin{equation}\label{eq: basic prop}
  m \cdot (\ell \circ r)= (m \cdot \ell)\circ r
\end{equation}
Let us consider functions of the form $f:L \rightarrow L$ and $g: L \rightarrow M$. We define the sets $F(L)=\{f| f:L \rightarrow L\}$ and $G(L,M)=\{g|g: L \rightarrow M\}$ which are simply denoted by $F$ and $G$ respectively. In $F$, $f_e$ denotes the identity function, i.e., $f_e(\ell)=\ell$ for any $\ell\in L$. In $G$, $g_e$ denotes the constant function $g_e(\ell)=1$ for every $\ell\in L$. Since composition of functions is associative then $(F,\circ, f_e)$ is a monoid. As the image of any function in $G$ is a subset of $M$, we also extend $\cdot$ in the following way: for any $g, g'\in G$, $g\cdot g'$ is defined as $(g\cdot g')(\ell)=g(\ell)\cdot g'(\ell)$ for any $\ell\in L$. Thus, $g\cdot g'$ is a well defined function of $G$. Clearly, $(G,\cdot, g_e)$ is a monoid.

Given the monoids $G$ and $F$ introduced above, we put our attention to the cartesian product $G\times F$ and pairs of functions $(g,f)\in G\times F$. We define the binary operation $\bullet :G \times F \rightarrow F$ as follows: given $g\in G$ and $f\in F$, $(g \bullet f)(\ell)= g(\ell)\cdot f(\ell)$ for any $\ell \in L$. Thus, $(g \bullet f)$ is a well defined function of $F$. 
Furthermore, for any $g, g' \in G$ and $f, f'\in F$, it is simple to prove that,
\begin{equation}\label{eq: associativity of f dot g}
(g\cdot g')\bullet f= g\bullet (g' \bullet f)
\end{equation}
\begin{equation}\label{eq: distributivity circ cdot}
(g \circ f')\bullet (f\circ f')= (g \bullet f)\circ f'
\end{equation}

We provide a product operation for elements of $G\times F$. The binary operation $\ast: (G\times F)^2 \rightarrow G\times F$ is defined as follows:
\begin{equation}\label{eq: right composition}
(g_1,f_1)\ast (g_2,f_2)=(g_1\cdot(g_2\circ f_1), f_2\circ f_1)
\end{equation} for any $(g_1,f_1)$, $(g_2,f_2)\in G\times F$. Obviously, $(g_1,f_1)\ast (g_2,f_2)\in G\times F$.
It is not difficult to prove that $\ast$ is associative. In addition, the pair $(g_e,f_e)$ is the identity element for $\ast$. Therefore, $(G\times F, \ast, (g_e,f_e))$ is a monoid. We abuse of the confidence of the reader by using the term \emph{composition} instead of $\ast$ when this operation is evident from its context of application.

Given a finite family $\{(g_i,f_i)\in G\times F\}_{i:1..n}$ with $n\geq 1$, the composition $(g_1,f_1)\ast (g_2,f_2)\ast...\ast (g_n,f_n)$, is denoted by $\ast|_{i=1}^n (g_i,f_i)$. By using (\ref{eq: right composition}) successively, the result can be expressed in the form:
\begin{equation}\label{eq: n-composition rigth}
\ast|_{i=1}^n (g_i,f_i)= (\prod_{i=1}^n g_i \circ (\circ|_{j=i-1}^1 f_j), \circ|_{i=n}^1 f_i)
\end{equation} where $\circ|$ and $\prod$ represent the quantifiers of $\circ$ and $\cdot$ respectively. By convention, $\circ|_{\emptyset}(.)=f_e$ and $\prod_{\emptyset}(.)=g_e$. Thus, (\ref{eq: n-composition rigth}) is also applicable when $n=0$. In that case, $\ast|_{\emptyset} (f_i,g_i)=(f_e,g_e)$.

\begin{example}\label{example 1}
As an example of (\ref{eq: n-composition rigth}), let us consider $n=3$:\\
\begin{eqnarray*}
\begin{array}{l}
   \ast|_{i=1}^3 (g_i,f_i)= (g_1, f_1)\ast (g_2,f_2)\ast (g_3,f_3)=\\
   ((g_1\circ f_e)\cdot (g_2\circ f_1)\cdot (g_3\circ(f_2\circ f_1)), f_3\circ f_2\circ f_1)=\\
   (g_1 \cdot (g_2\circ f_1)\cdot (g_3\circ(f_2\circ f_1)),f_3\circ f_2\circ f_1)
\end{array}
\end{eqnarray*} \hfill \qed
\end{example}

In order to obtain a more compact notation, we will write equation (\ref{eq: n-composition rigth}) in the form
\begin{equation}\label{eq: compact n-composition rigth}
[(g_i,f_i)]_1^n= ([g_i \circ [f_j]_{i-1}^1]_1^n, [f_i]_n^1)
\end{equation} where $[(g_i,f_i)]_1^n=\ast|_{i=1}^n (g_i,f_i)$, $[f_i]_n^1= \circ|_{i=n}^1 f_i$, and $[g_i \circ [f_j]_{i-1}^1]_1^n= \prod_{i=1}^n g_i \circ (\circ|_{j=i-1}^1 f_j)$. By using this notation, we obtain, by (\ref{eq: right composition}), that
\begin{equation}\label{eq: recurrence of composition}
\begin{array}{l}
  [(g_i,f_i)]_1^{n+1}= [(g_i,f_i)]_1^{n} \ast (g_{n+1},f_{n+1}) = \\
  ([g_i \circ [f_j]_{i-1}^1]_1^n \cdot(g_{n+1} \circ [f_j]_{n}^1), f_{n+1} \circ [f_i]_n^1)
\end{array}
\end{equation} for the composition of a family of $n+1$ pairs of functions in $G\times F$, with $n\geq 0$.

\begin{remark}\label{rem: for induction}
Let us observe that, by (\ref{eq: compact n-composition rigth}), $[(g_i,f_i)]_1^{n+1}=([g_i \circ [f_j]_{i-1}^1]_1^{n+1}, [f_i]_{n+1}^1)$. In some proofs, we are interested in the operation
$[g_i \circ [f_j]_{i-1}^1]_1^{n+1} \bullet [f_i]_{n+1}^1$.
\begin{eqnarray*}
\begin{array}{l}
[g_i \circ [f_j]_{i-1}^1]_1^{n+1} \bullet [f_i]_{n+1}^1 = (\mbox{\emph{by (\ref{eq: recurrence of composition})}}) =\\
([g_i \circ [f_j]_{i-1}^1]_1^n \cdot(g_{n+1} \circ [f_j]_{n}^1))\bullet (f_{n+1} \circ [f_i]_n^1) = (\mbox{\emph{by (\ref{eq: associativity of f dot g})}}) =\\
([g_i \circ [f_j]_{i-1}^1]_1^n) \bullet ((g_{n+1} \circ [f_j]_{n}^1) \bullet (f_{n+1} \circ [f_i]_n^1))= (\mbox{\emph{by (\ref{eq: distributivity circ cdot})}} )=\\
([g_i \circ [f_j]_{i-1}^1]_1^n)\bullet ((g_{n+1} \bullet f_{n+1})\circ [f_i]_n^1)
\end{array}
\end{eqnarray*} In conclusion,
\begin{equation}\label{eq: induction for n+1}
 [g_i \circ [f_j]_{i-1}^1]_1^{n+1} \bullet [f_i]_{n+1}^1 = [g_i \circ [f_j]_{i-1}^1]_1^n\bullet ((g_{n+1} \bullet f_{n+1})\circ [f_i]_n^1)
\end{equation}
\end{remark}\hfill \qed

For each word-transformation $r:\Sast \rightarrow \Sast$, we define the function $\lder{r}\in F$, $\lder{r}:L \rightarrow L$, as $\lder{r}(\ell)=\ell \circ r$ for any $\ell \in L$. We say that $\lder{r}(\ell)$ is the \emph{derivative of} $\ell$ by the word-transformation $r$.

\begin{example}\label{example 2}
Let us consider, for each $\alpha \in \Sast$, the transformation $\underline{\alpha}:\Sast \rightarrow \Sast$ defined by $\underline{\alpha}(\gamma)= \alpha\gamma$ for any $\gamma \in \Sast$. Thus, the derivative $\lder{\underline{\alpha}}(\ell)$, satisfies that $\lder{\underline{\alpha}}(\ell)= \ell \circ \underline{\alpha}$ by the definition given above. Then, $\lder{\underline{\alpha}}(\ell)(\gamma)= \ell(\underline{\alpha}(\gamma))=\ell(\alpha\gamma)$ for each word $\gamma$. Let us observe that $\lder{\underline{\alpha}}(\ell)$ may be viewed as a generalization of the Brzozowski derivative of an ordinary language by a word \cite{Brzozowski1962}. In addition, $\lder{\underline{\beta}}\circ\lder{\underline{\alpha}}=\lder{\underline{\alpha} \circ \underline{\beta}}=\lder{\underline{\alpha\beta}}$ holds for any words $\alpha$ and $\beta$. Clearly, $\lder{\underline{\varepsilon}}=f_e$. In the rest of this paper, $\lder{\underline{\alpha}}$ is simply denoted by  $\lder{\alpha}$ for any word $\alpha$. \hfill \qed


\end{example}

Given a word-transformation $r:\Sast \rightarrow \Sast$ and a pair $(g,f)\in G\times F$, the next equation holds:

\begin{equation}\label{eq: g dot (delta f)}
  g\bullet(\lder{r}\circ f)= \lder{r}\circ(g\bullet f)
\end{equation}
\noindent \emph{Proof}: For any $\ell \in L$,
\begin{eqnarray*}
\begin{array}{l}
  (g\bullet(\lder{r}\circ f))(\ell)= g(\ell)\cdot(\lder{r}(f(\ell)))= g(\ell)\cdot(f(\ell)\circ r)= (\mbox{ by (\ref{eq: basic prop}) })=\\
  (g(\ell) \cdot f(\ell))\circ r = (g\bullet f)(\ell)\circ r= \lder{r}((g\bullet f)(\ell))= (\lder{r} \circ (g\bullet f))(\ell)
\end{array}
\end{eqnarray*}\hfill \qed

\section{Factorizations on $L(\Sast,M)$}\label{sec: factorizations}

Let us observe that the identity element $(g_e,f_e)\in G\times F$ satisfies that $g_e \bullet f_e= f_e$ since, for any $\ell \in L$, $(g_e \bullet f_e)(\ell)= g_e(\ell)\cdot f_e(\ell)= 1 \cdot \ell=\ell$. We assume the hypothesis that there exist other pairs $(g,f)\in G\times F$ with the same property, i.e., $(g \bullet f)(\ell)= g(\ell)\cdot f(\ell)=\ell$ for any $\ell\in L$. In that case, $g(\ell)\in M$ divides each value $\ell(\alpha)\in M$ for any word $\alpha$, i.e., it is a common factor for $\ell$. Thus, we may say that the pair $(g,f)$ factorizes $L$. In general,
\begin{definition}\label{def: factorization}
  A pair of functions $(g,f)\in G\times F$ is a \emph{factorization} on $L$ if $(g,f)$ satisfies that
  \begin{equation}\label{eq: factorization}
  g \bullet f= f_e
  \end{equation}
\end{definition}
The identity element $(g_e,f_e)$ is called the \emph{trivial factorization} on $L$. We will study the properties obtained from the definition of factorization without considering additional properties about the monoid or the functions involved in the factorization. If $(g,f)$ is a factorization on $L$ then
\begin{equation}\label{eq: basic prop fact1}
 \lder{r}= \lder{r}\circ (g\bullet f)= g \bullet (\lder{r}\circ f)
\end{equation}
\begin{equation}\label{eq: basic prop fact2}
  \lder{r}= (g \bullet f)\circ \lder{r}= (g\circ \lder{r})\bullet (f\circ \lder{r})
\end{equation} for any word-transformation $r:\Sast\rightarrow \Sast$. Those results are consequence of (\ref{eq: g dot (delta f)}), (\ref{eq: distributivity circ cdot}), and Definition \ref{def: factorization}.

\begin{lemma}\label{lem: composition of factorizations}
 Let $\{(g_i,f_i)\in G\times F\}_{i:1..n}$ be an arbitrary finite family of $n\geq 0$ factorizations on $L$. The composition $[(g_i,f_i)]_1^n$ is a factorization on $L$.
\end{lemma}
\noindent \emph{Proof}: By (\ref{eq: compact n-composition rigth}), $[(g_i,f_i)]_1^n= ([g_i \circ [f_j]_{i-1}^1]_1^n, [f_i]_n^1)$. By induction on $n$:\\

\noindent - Basis. if $n=0$ then $[(g_i,f_i)]_{\emptyset}= (g_e,f_e)$, the trivial factorization.\\
\noindent - Hypothesis. Let us assume that $[(g_i,f_i)]_1^n$ is a factorization on $L$ for some arbitrary $n\geq0$, i.e., $[g_i \circ [f_j]_{i-1}^1]_1^n \bullet [f_i]_n^1= f_e$ (Definition \ref{def: factorization}).\\
\noindent - Induction Step. Let us consider the composition $[(g_i,f_i)]_1^{n+1}$ where $(g_{n+1},f_{n+1})$ is a factorization on $L$. By (\ref{eq: induction for n+1}), the fact that $g_{n+1} \bullet f_{n+1}= f_e$, and induction Hypothesis, $[g_i \circ [f_j]_{i-1}^1]_1^{n+1} \bullet [f_i]_{n+1}^1=f_e$. Therefore, $[(g_i,f_i)]_1^{n+1}$ is a factorization on $L$. \hfill \qed

\begin{lemma}\label{lemma: contraction}
Let $\{(g'_i,f'_i)\in G\times F\}_{i:1..n}$ be a finite family of $n\geq 0$ factorizations on $L$. Let $\{r_i:\Sast \rightarrow \Sast\}_{i:1..n}$ be a finite family of $n\geq 0$ word-transformations. For each $i:1..n$, the pair $(g_i,f_i)\in G\times F$ is defined as $(g_i,f_i)= (g'_i\circ \lder{r_i}, f'_i \circ \lder{r_i})$ where $\lder{r_i}$ is the derivative operator by the word-transformation $r_i$. The composition of the family $\{(g_i,f_i)\in G\times F\}_{i:1..n}$, $[(g_i,f_i)]_1^n= ([g_i \circ [f_j]_{i-1}^1]_1^n, [f_i]_n^1)$, satisfies that
\begin{equation}\label{eq: contraction of subfactorizations}
[g_i \circ [f_j]_{i-1}^1]_1^n \bullet [f_i]_n^1= \lder{\circ|_{i=1}^n r_i}
\end{equation}
\end{lemma}
\noindent \emph{Proof}: Let us recall that $\lder{r_i}\in F$ and $\lder{r_i}(\ell)=\ell \circ r_i$ for any $\ell \in L$. By induction on $n$:\\

\noindent - Basis. if $n=0$ then $[(g_i,f_i)]_{\emptyset}= (g_e,f_e)$. By convention $\circ|_{\emptyset} r_i$ is the identity word-transformation. Thus, $\lder{\circ|_{\emptyset} r_i}=f_e$. Therefore, $g_e\bullet f_e= f_e$ since $(g_e,f_e)$ is the trivial factorization on $L$.\\
\noindent - Hypothesis. Let us assume that (\ref{eq: contraction of subfactorizations}) holds for an arbitrary $n\geq 0$.\\
\noindent - Induction Step. Let us consider the composition $[(g_i,f_i)]_1^{n+1}$ where each $(g_i,f_i)$ is defined in Lemma \ref{lemma: contraction}. The new pair $(g_{n+1}, f_{n+1})$ is also defined as $(g_{n+1},f_{n+1})= (g'_{n+1}\circ \lder{r_{n+1}}, f'_{n+1} \circ \lder{r_{n+1}})$ for a given factorization on $L$, $(g'_{n+1},f'_{n+1})$, and a word-transformation $r_{n+1}$.\\
\noindent By definition of $(g_{n+1}, f_{n+1})$, $g_{n+1}\bullet f_{n+1}= (g'_{n+1}\circ \lder{r_{n+1}}) \bullet (f'_{n+1} \circ \lder{r_{n+1}})$.\\
\noindent Then, by (\ref{eq: basic prop fact2}), $g_{n+1}\bullet f_{n+1}= \lder{r_{n+1}}$.\\
\noindent By (\ref{eq: induction for n+1}), and substitution,\\
\noindent $[g_i \circ [f_j]_{i-1}^1]_1^{n+1} \bullet [f_i]_{n+1}^1 = [g_i \circ [f_j]_{i-1}^1]_1^n\bullet (\lder{r_{n+1}}\circ [f_i]_n^1)$.\\
\noindent By (\ref{eq: g dot (delta f)}), $[g_i \circ [f_j]_{i-1}^1]_1^{n+1} \bullet [f_i]_{n+1}^1= \lder{r_{n+1}} \circ ([g_i \circ [f_j]_{i-1}^1]_1^n \bullet [f_i]_n^1)$.\\
\noindent By Hypothesis and definition of $\lder{r_{n+1}}$,\\
\noindent $[g_i \circ [f_j]_{i-1}^1]_1^{n+1} \bullet [f_i]_{n+1}^1 = \lder{r_{n+1}} \circ \lder{\circ|_{i=1}^n r_i}= \lder{\circ|_{i=1}^{n+1} r_i}$. Therefore, the Lemma holds. \hfill \qed\\

Previous results have been provided for arbitrary factorizations, word-transformations and their derivatives. The importance of factorizations on $L$ and the equation (\ref{eq: contraction of subfactorizations}) is that they provide a reasonable theoretical basis for recognizability of $M$-languages via deterministic finite automata with values on a monoid $M$ as we will show in the next sections.

\section{Deterministic finite automata with values on a monoid}\label{sec: M-dfa}

We present a short introduction to deterministic finite automata with values on a monoid.

\begin{definition}\label{def: M-dfa} Let $(M, \cdot, 1)$ be a monoid. A Deterministic Finite Automaton with values on the monoid $M$, ($M$-DFA in short), is a tuple $A= (Q, \Sigma, u, i_u, \delta, w, \rho)$ where
\begin{itemize}
  \item $Q$ is a finite nonempty set of states;
  \item $\Sigma$ is a finite alphabet;
  \item $u\in Q$ is the unique \emph{initial state};
  \item $i_u \in M$ is the \emph{initial value} assigned to the initial state $u\in Q$;
  \item $\delta: Q \times \Sigma \rightarrow Q$ is the \emph{state-transition function};
  \item $w: Q \times \Sigma \rightarrow M$ is the \emph{monoid-transition function} that assigns values from $M$ to each transition; and
  \item $\rho: Q \rightarrow M$ is the \emph{final-function} that assigns \emph{final values} from $M$ to each state in $Q$.
\end{itemize}
\end{definition}

The state-transition function $\delta$ is extended to $\Sast$. The extended function $\delta^\ast:Q\times \Sast\rightarrow Q$ is defined as
\begin{equation}\label{eq: delta_ast}
  \mbox{(i) } \delta^\ast(q,\varepsilon)= q; \mbox{ and, } \mbox{(ii) } \delta^\ast(q, \alpha\sigma)= \delta(\delta^\ast(q,\alpha), \sigma)
\end{equation} for any $q\in Q$, $\alpha\in \Sast$, and $\sigma\in \Sigma$. As $\delta$ is a total function then $\delta^\ast(q,\alpha)\in Q$ for every word $\alpha$. Due this fact, it is common to say that $A$ is \emph{complete}. In the rest of this paper, $\delta^\ast(q,\alpha)$ is simply denoted $q\alpha$ for each state $q$ of $A$, and word $\alpha$. As $A$ is a deterministic automaton, $q\alpha$ is the unique \emph{reachable state} from $q$ by the word $\alpha$. In other words, $q\alpha$ is an \emph{accessible state} from $q$ (by the word $\alpha$). An $M$-DFA $A$ is \emph{accessible} if $Q= \{u\alpha | \alpha \in \Sast\}$, i.e., any state in $Q$ is accessible from the initial state.

The monoid-transition function $w$ is also extended to $\Sast$. The extended function $w^\ast: Q\times \Sast\rightarrow M$ is defined as
\begin{equation}\label{eq: omega_ast}
  \mbox{(i) } w^\ast(q,\varepsilon)= 1; \mbox{ and, } \mbox{(ii) } w^\ast(q, \alpha\sigma)= w^\ast(q,\alpha) \cdot w(q\alpha, \sigma)
\end{equation} for any $q\in Q$, $\alpha\in \Sast$, and $\sigma\in \Sigma$. Let us observe that the definition of $w^\ast$ uses the extended function $\delta^\ast$.

Let $\alpha\in \Sast$ be a word. The length of $\alpha$ is denoted by $|\alpha|$. The $k$-th prefix of $\alpha$ is $\alpha[k]$ where $0\leq k\leq |\alpha|$. By convention, $\alpha[0]=\varepsilon$. In addition, the $k$-th symbol in $\alpha$ is denoted by $\alpha(k)$ where, by convention, $\alpha(k)=\varepsilon$ when $k<1$ or $k>|\alpha|$. Taken into account such a notation, by (\ref{eq: omega_ast}), we can expand $w^\ast(q,\alpha)$ as follows:
\begin{equation}\label{eq: omega_ast of alpha}
\begin{array}{l}
  w^\ast(q,\alpha)= w(q, \alpha(1))\cdot ...w(q\alpha[i-1],\alpha(i))...\cdot w(q\alpha[n-1],\alpha(n))\\
  \\
   \displaystyle {= \prod_{i=1}^{n} w(q\alpha[i-1],\alpha(i))}
\end{array}
\end{equation} for any word $\alpha$ of length $n\geq 0$ and $q\in Q$. It is also clear that, for any two words $\alpha$ and $\beta$, and $q\in Q$,
\begin{equation}\label{eq: omega_ast of alpha and beta}
  w^\ast(q,\alpha\beta)= w^\ast(q,\alpha) \cdot w^\ast(q\alpha, \beta)
\end{equation}

Given those previous definitions and extended functions, it is possible to define the $M$-language recognized (or generated) by an $M$-DFA. Let $A= (Q, \Sigma, u, i_u, \delta, w, \rho)$ be an $M$-DFA. The $M$-language \emph{recognized} by $A$, denoted $\mathcal{A}$, $\mathcal{A}\in L$, is defined by
\begin{equation}\label{eq: Mlanguage of A}
  \mathcal{A}(\alpha)= i_u \cdot w^\ast(u,\alpha) \cdot \rho(u\alpha)
\end{equation} for any $\alpha \in \Sast$.
In addition, for each state $q$ of $A$, the $M$-language $\mathcal{A}_q \in L$ is defined by
\begin{equation}\label{eq: Mlanguage of Aq}
  \mathcal{A}_q(\alpha)= w^\ast(q,\alpha) \cdot \rho(q\alpha)
\end{equation} for any $\alpha \in \Sast$. For each word $\alpha\in \Sast$ and its derivative $\lder{\alpha}\in F$ (see Example \ref{example 2}), it is simple to prove that
\begin{equation}\label{eq: Mlanguage derivative alpha Aq}
  \lder{\alpha}(\mathcal{A}_q)= w^\ast(q,\alpha)\cdot \mathcal{A}_{q\alpha}
\end{equation} for any $q\in Q$. Finally, $\lder{\alpha}(\mathcal{A})= i_u \cdot \lder{\alpha}(\mathcal{A}_u)$.

\begin{definition}\label{def: recognizable}
An $M$-language $\ell \in L$, is a \emph{recognizable} $M$-language, or simply \emph{recognizable}, if $\ell$ is recognized by some $M$-DFA.
\end{definition}
Two trivial consequences of Definition \ref{def: recognizable} are:
\begin{quote}
({\bf Rcg1}) If $\ell \in L$ is recognizable then, for any $m\in M$, $m\cdot \ell$ is recognizable; and\\
({\bf Rcg2}) if $\ell \in L$ is recognizable then, there exists $\ell' \in L$ such that $\ell'$ is recognizable and $\ell= m\cdot \ell'$ for some $m\in M$.
\end{quote}
If $\ell\in L$ is recognized by $A=(Q,\Sigma, u, i_u, \delta, w, \rho)$, i.e., $\mathcal{A}= \ell$, then, by (\ref{eq: Mlanguage of A}), ({\bf Rcg1}) $m\cdot \ell$ is recognized by the automaton $A'=(Q,\Sigma, u, m \cdot i_u, \delta, w, \rho)$. For ({\bf Rcg2}), let us consider $A_u=(Q,\Sigma, u, 1, \delta, w, \rho)$. Then,  by (\ref{eq: Mlanguage of A}), $\ell'= \mathcal{A}_u$ and $m= i_u$.

Given two $M$-DFAs, $A$ and $B$, we say that $A$ is (language) \emph{equivalent} to $B$ when they recognize the same $M$-language, i.e., $\mathcal{A}=\mathcal{B}$.

Among the equivalent automata to $A$, we may find a \emph{minimal} one. The minimal automaton has the minimal number of states for recognizing the same $M$-language.
\begin{property}\label{prop: necessary cond min}
Let $A=(Q,\Sigma, u, i_u, \delta, w, \rho)$ be an $M$-DFA. If $A$ is minimal then $A$ satisfies the next conditions:
\begin{enumerate}
\item $A$ is an accessible $M$-DFA.
\item ({\bf NcndS}) For all states $p,q \in Q$,\\
($\exists m\in M: m\cdot \mathcal{A}_p=\mathcal{A}_q \vee \mathcal{A}_p=m \cdot \mathcal{A}_q) \Rightarrow p=q$
\item ({\bf NcndW}) For all states $p,q \in Q$, $\mathcal{A}_p= \mathcal{A}_q \Rightarrow p=q$
\end{enumerate}
\end{property}
\noindent \emph{Proof}: Let $A=(Q,\Sigma, u, i_u, \delta, w, \rho)$ be a minimal $M$-DFA.\\
\noindent 1. If $A$ is not accessible then there is an equivalent $M$-DFA $A'$ with a lesser number of states than $A$. $A'$ is built by removing the unaccessible states of $A$. This is in contradiction with the initial hypothesis.\\

\noindent 2. Let us consider that for two states $p\neq q$, $m\cdot \mathcal{A}_p= \mathcal{A}_q$ for some $m\in M$. Let us define the $M$-DFA $A'= (Q, \Sigma, u, i_u, \delta', w', \rho)$. $A'$ is exactly equal to $A$ excepting that, for any $s\in Q$ and $\sigma \in \Sigma$, if $\delta(s,\sigma)=q$ then $\delta'(s,\sigma)=p$ and $w'(s,\sigma)=w(s,\sigma)\cdot m$. That is, every transition $(s,\sigma)$ ending in $q$ in $A$ ends in $p$ in $A'$.
It is easily to prove that $A'$ is equivalent to $A$. However, in $A'$, $q$ is not an accessible state. Removing $q$ in $A'$, constructs an equivalent automaton to $A$ but with less states. Again, this is a contradiction.\\

\noindent 3. {\bf NcndW} is consequence of {\bf NcndS} when $m= 1$. \hfill \qed\\

Those previous conditions are \emph{necessary} conditions for an $M$-DFA to be a minimal one. Let us observe that ({\bf NcndS}) is stronger than ({\bf NcndW}). It is also possible to provide a \emph{sufficient} condition for minimality.
\begin{property}\label{prop: suficcient cond min}
Let $A=(Q,\Sigma, u, i_u, \delta, w, \rho)$ be an accessible $M$-DFA, then\\
\begin{equation}
  \begin{array}{l}
   \neg(\exists \alpha,\beta\in \Sast, \ell\in L, m,m'\in M:\\
   u\alpha\neq u\beta \wedge \lder{\alpha}(\mathcal{A})= m\cdot \ell \wedge \lder{\beta}(\mathcal{A})= m' \cdot \ell)\\
    \Rightarrow \\
    A \mbox{ is minimal}
  \end{array}
\end{equation}
\end{property}
\noindent \emph{Proof}: As $A$ is accessible, $Q=\{u\alpha~| \alpha\in \Sast\}$. Assume that $\|Q\|= n$. As $A$ is a deterministic automaton, there exists $n$ different words, $\alpha_1$...$\alpha_n$, such that $Q=\{u\alpha_i~| i:1..n\}$. Suppose that $A$ is not minimal. Then, there exists a minimal $M$-DFA $A'=(Q', \Sigma, u', i_{u'}, \delta', w',\rho')$ equivalent to $A$ with $\|Q'\|< \|Q\|$. As $A'$ is complete $u'\alpha_i\in Q'$ for each $i:1..n$. By the Pigeonhole principle, there are at least two different words $\alpha_k$ and $\alpha_{k'}$, with $1\leq k < k'\leq n$, such that $u_{\alpha_k}\neq u_{\alpha_{k'}}$ and $u'_{\alpha_k}= u'_{\alpha_{k'}}$. Call $\alpha_k$ and $\alpha_{k'}$ by $\alpha$ and $\beta$ respectively. Thus, $\lder{\alpha}(\mathcal{A})=\lder{\alpha}(\mathcal{A'})$ and  $\lder{\beta}(\mathcal{A})=\lder{\beta}(\mathcal{A'})$ since $A$ is equivalent to $A'$. By (\ref{eq: Mlanguage derivative alpha Aq}), $\lder{\alpha}(\mathcal{A'})= i_{u'}\cdot w'^\ast(u',\alpha)\cdot \mathcal{A'}_{u'\alpha}$ and $\lder{\beta}(\mathcal{A'})= i_{u'}\cdot w'^\ast(u',\beta)\cdot \mathcal{A'}_{u'\beta}$. As $u'_{\alpha}= u'_{\beta}$, then $\mathcal{A'}_{u'\alpha}=\mathcal{A'}_{u'\beta}=\ell$. In conclusion, $u\alpha\neq u\beta \wedge \lder{\alpha}(\mathcal{A})= m\cdot \ell \wedge \lder{\beta}(\mathcal{A})= m' \cdot \ell$. A contradiction happens and the property holds. \hfill \qed\\

In general, we say that an $M$-DFA $A=(Q,\Sigma, u, i_u, \delta, \omega, \rho)$ is \emph{transition-equalized} if for any $p$, $q\in Q$ and $\sigma$, $\tau \in \Sigma$,
\begin{equation}\label{eq: transition-equalized automaton}
\lder{\sigma}(\mathcal{A}_q)=\lder{\tau}(\mathcal{A}_p) \Rightarrow \delta(q,\sigma)= \delta(p,\tau) \wedge w(q,\sigma)=w(p,\tau)
\end{equation}

This property has an important relation with the construction of factorizations induced by an $M$-DFA (see section \ref{sec: natural factorizations}).

\section{Recognizability via factorizations on $L(\Sast,M)$}\label{sec: recognizability}

In this section, our main result states a \emph{sufficient condition} for the recognizability of $M$-languages via factorizations on $L$. We recall that, for a word $\alpha\in \Sast$, $\alpha(i)$ denotes the $i$-th symbol in $\alpha$ and $\alpha[i]$ denotes the $i$-th prefix of $\alpha$. Given a factorization on $L$, $(g,f)$, and a word $\alpha$, let us define the pairs $(g\circ \lder{\alpha(i)}, f\circ \lder{\alpha(i)})\in G \times F$ with $i: 1.. |\alpha|$, and the composition $[(g\circ \lder{\alpha(i)}, f\circ \lder{\alpha(i)})]_1^{|\alpha|}$. By (\ref{eq: n-composition rigth}) and the notation given in (\ref{eq: compact n-composition rigth}), that composition returns the pair $(W^{(g,f)}_\alpha,S^{(g,f)}_\alpha)\in G \times F$, where
\begin{equation}\label{eq: def of S(g,f)alpha}
S^{(g,f)}_\alpha= [f\circ \lder{\alpha(i)}]_{|\alpha|}^1
\end{equation}
\begin{equation}\label{eq: def of W(g,f)alpha}
  W^{(g,f)}_\alpha= [g \circ \lder{\alpha(i)}\circ S^{(g,f)}_{\alpha[i-1]}]_1^{{|\alpha|}}
\end{equation}

We write $S_\alpha(\ell)$ and $W_\alpha(\ell)$ instead of $S^{(g,f)}_\alpha$ and $W^{(g,f)}_\alpha$ in many part of the text when the factorization $(g,f)$ is clear in the context of discussion. We recognize that it is not simple to provide an interpretations of these functions. This is the reason for providing their main properties:
\begin{eqnarray}\label{prop1}
 S_\varepsilon = f_e \mbox{ and } W_\varepsilon = g_e \\ \label{prop1b}
 S_\sigma= f\circ \lder{\sigma} \mbox{ and } W_\sigma= g \circ \lder{\sigma}\\ \label{prop2}
 S_\beta \circ S_\alpha= S_{\alpha\beta} \\ \label{prop3}
 W_\alpha \cdot (W_\beta \circ S_\alpha) = W_{\alpha\beta}\\ \label{prop4}
 W_\alpha \bullet S_\alpha = \lder{\alpha}\\ \label{prop5}
 (W_\beta \circ S_\alpha)\bullet (S_\beta \circ S_\alpha)= \lder{\beta}\circ S_\alpha
\end{eqnarray} for any $\sigma\in \Sigma$, and $\alpha$, $\beta\in \Sast$.\\

Equations (\ref{prop1}), (\ref{prop1b}), (\ref{prop2}) and (\ref{prop3}) are directly obtained from the definition of $S_\alpha$ (\ref{eq: def of S(g,f)alpha}) and $W_\alpha$ (\ref{eq: def of W(g,f)alpha}). Equation (\ref{prop4}) is a consequence of the fact that $(W_\alpha, S_\alpha)$ is an element of $G\times F$ that satisfies (\ref{eq: contraction of subfactorizations}) in Lemma \ref{lemma: contraction}. The last equation (\ref{prop5}) is obtained by (\ref{eq: distributivity circ cdot}) and the previous one (\ref{prop4}).

The main justification for introducing such a pair $(W_\alpha, S_\alpha)$ is that, for any $\ell \in L$ and $\alpha \in \Sast$,
\begin{eqnarray*}
  (W_\alpha \bullet S_\alpha)(\ell)= \lder{\alpha}(\ell) \\
   W_\alpha(\ell) \cdot S_\alpha(\ell)= \ell \circ \underline{\alpha}\\
  (W_\alpha(\ell) \cdot S_\alpha(\ell))(\varepsilon)= (\ell \circ \underline{\alpha})(\varepsilon)\\
   W_\alpha(\ell) \cdot (S_\alpha(\ell))(\varepsilon)= \ell(\alpha)
\end{eqnarray*} In conclusion, for any factorization on $L$, $(g,f)$, $\ell \in L$ and $\alpha \in \Sast$:
\begin{equation}\label{eq: justification}
  W^{(g,f)}_\alpha(\ell) \cdot (S^{(g,f)}_\alpha(\ell))(\varepsilon)= \ell(\alpha)
\end{equation}
This last equation suggests us a way for constructing the $M$-language $\ell$ taken into account that, by (\ref{prop3}), $W^{(g,f)}_\alpha(\ell)$, with $|\alpha|=n$, can be expanded as
\begin{equation}\label{eq: Walpha expanded}
\begin{array}{l}
 W_\alpha(\ell)= W_{\alpha(1)}(\ell) \cdot (W_{\alpha(2)}\circ S_{\alpha[1]})(\ell)\cdot ...\cdot (W_{\alpha(n)}\circ S_{\alpha[n-1]})(\ell) \\
  W_\alpha(\ell)= \prod_{i=1}^{|\alpha|}(W_{\alpha(i)}\circ S_{\alpha[i-1]})(\ell)
\end{array}
\end{equation}
By (\ref{eq: justification}) and the considered expansion, each value $\ell(\alpha)\in M$, is the product of $|\alpha|+1$ values from $M$ for any factorization. The reader may compare the language accepted by an $M$-DFA (\ref{eq: Mlanguage of A}) with (\ref{eq: justification}), and (\ref{eq: omega_ast of alpha}) with (\ref{eq: Walpha expanded}). The main question is how to ensure that every $\ell(\alpha)$ is finitely generated by some automaton given a factorization. This question is solved in the following lemma for recognizability.

\begin{lemma}\label{theo: recognizability}
  Let $\ell\in L$ be an $M$-language. If there exists a factorization on $L$, $(g,f)$, such that the set $\{S^{(g,f)}_\alpha(\ell) | \alpha \in \Sast\}$ is finite, then $\ell$ is a recognizable $M$-language.
\end{lemma}
\noindent \emph{Proof}: We construct an $M$-DFA that recognizes $\ell \in L$. By assumption $\{S^{(g,f)}_\alpha(\ell) | \alpha \in \Sast\}$ is finite. It is a nonempty set since $\ell$ is in that set. In the rest of the proof, we omit the superscript $(g,f)$ excepting for some definitions. Let us consider the following automaton:
\begin{definition}\label{def: nerode automaton} $N^{(g,f)}_{\ell}=(Q, \Sigma, S_{\varepsilon}(\ell), 1, \delta, w, \rho)$:
\begin{itemize}
\item $Q= \{S_\alpha(\ell) | \alpha \in \Sast\}$ is the set of sates, each state is an $M$-language;
\item the initial state is $S_{\varepsilon}(\ell) \in Q$, by (\ref{prop1}), $S_{\varepsilon}(\ell)=\ell$;
\item the initial value of $S_{\varepsilon}(\ell)$ is $1$;
\item the state-transition function $\delta$ is defined as\\
      $\delta(S_\alpha(\ell), \sigma)=S_{\alpha\sigma}(\ell)$ for any $\alpha\in \Sast$ and $\sigma \in \Sigma$;
\item the monoid-transition function $w$ is defined as\\
      $w(S_\alpha(\ell), \sigma)= (W_\sigma \circ S_\alpha)(\ell)$ for any $\alpha\in \Sast$ and $\sigma \in \Sigma$; and
\item for each state, its final value is $\rho(S_\alpha(\ell))= (S_\alpha(\ell))(\varepsilon)$ for any $\alpha\in \Sast$.
\end{itemize}
\end{definition}

\noindent Both $\delta:Q\times \Sigma\rightarrow Q$ and $w:Q\times \Sigma\rightarrow M$ are well defined functions. Given two words $\alpha$ and $\beta$, if $S_\alpha(\ell)=S_\beta(\ell)$ then $S_{\alpha\sigma}(\ell)= (S_\sigma \circ S_\alpha)(\ell)= (S_\sigma \circ S_\beta)(\ell)= S_{\beta\sigma}(\ell)$; and, trivially, $(W_\sigma \circ S_\alpha)(\ell)= (W_\sigma \circ S_\beta)(\ell)$. Therefore, the automaton $N^{(g,f)}_\ell$ is a well defined $M$-DFA since $Q$ is a nonempty finite set by the conditions of the lemma.\\

\noindent It is simple to show that $\delta^\ast(S_\alpha(\ell), \beta)= S_{\alpha\beta}(\ell)$ for any words $\alpha$ and $\beta$. By notation, $\delta^\ast(S_\alpha(\ell), \beta)$ is represented as $S_{\alpha}(\ell)\beta$ (as in section \ref{sec: M-dfa}). Thus, $S_{\alpha}(\ell)\beta= S_{\alpha\beta}(\ell)$.\\

\noindent In the following, we prove that the $M$-language recognized by $N^{(g,f)}_\ell$, in notation $\mathcal{N}^{(g,f)}_\ell$, satisfies $\mathcal{N}^{(g,f)}_\ell=\ell$.\\

\noindent For any $\alpha\in \Sast$:\\
\noindent By (\ref{eq: Mlanguage of A}), $\mathcal{N}^{(g,f)}_\ell(\alpha)= 1 \cdot w^\ast(S_\varepsilon(\ell),\alpha)\cdot \rho(S_\varepsilon(\ell)\alpha)$.\\
\noindent Since, $S_\varepsilon(\ell)\alpha= S_{\alpha}(\ell)$, then $ \mathcal{N}^{(g,f)}_\ell(\alpha) = w^\ast(S_\varepsilon(\ell),\alpha)\cdot \rho(S_{\alpha}(\ell))$,\\
\noindent By (\ref{eq: omega_ast of alpha}), $\mathcal{N}^{(g,f)}_\ell(\alpha) = (\prod_{i=1}^{|\alpha|}w(S_\varepsilon(\ell)\alpha[i-1], \alpha(i))) \cdot \rho(S_{\alpha}(\ell))$, and again, $S_\varepsilon(\ell)\alpha[i-1]= S_{\alpha[i-1]}(\ell)$, then, $\mathcal{N}^{(g,f)}_\ell(\alpha) = (\prod_{i=1}^{|\alpha|}w(S_{\alpha[i-1]}(\ell), \alpha(i))) \cdot \rho(S_{\alpha}(\ell))$\\
\noindent By definition of $w()$ and $\rho()$ given for the automaton $N^{(g,f)}_\ell$,\\
\noindent $\mathcal{N}^{(g,f)}_\ell(\alpha) = (\prod_{i=1}^{|\alpha|} (W_{\alpha(i)} \circ S_{\alpha[i-1]})(\ell))\cdot (S_{\alpha}(\ell))(\varepsilon)$\\
\noindent By the considered expansion of $W_\alpha(\ell)$ given in (\ref{eq: Walpha expanded}),\\
\noindent $\mathcal{N}^{(g,f)}_\ell(\alpha)= W_\alpha(\ell) \cdot (S_\alpha(\ell))(\varepsilon)$.\\
\noindent Finally, by (\ref{eq: justification}), $\mathcal{N}^{(g,f)}_\ell(\alpha)= \ell(\alpha)$.\\

Therefore, $\mathcal{N}^{(g,f)}_\ell=\ell$. This fact concludes that $\ell$ is a recognizable $M$-language via the factorization $(g,f)$. \hfill \qed\\

As Lemma \ref{theo: recognizability} indicates, $\ell\in L$ is recognizable if the automaton $N^{(g,f)}_\ell$ is an $M$-DFA for some factorization on $L$, $(g,f)$. 
In the construction provided in Definition \ref{def: nerode automaton}, $N^{(g,f)}_\ell$ depends on $\ell$ and the factorization $(g,f)$. The initial value of this automaton may be also changed without modifying the rest of components. In that case, we write $N^{(g,f)}(\ell, 1)$ for the original automaton and $N^{(g,f)}(\ell, m)$ for the same automaton but with initial value $m\in M$.

Let us observe that the finiteness property of the set of states $Q^{(g,f)}_\ell= \{S_\alpha(\ell) | \alpha \in \Sast\}$ of $N^{(g,f)}(\ell, 1)$, is due to the fact that infinitely many words satisfy $S_\alpha(\ell)= S_\beta(\ell)$. This argument allows us to propose a \emph {right congruence} on $\Sast$.
\begin{definition}\label{def: right congruence} Let $(g,f)$ be a factorization on $L$ and let $\ell$ be an $M$-language. The binary relation on $\Sast$ denoted by $\equiv^{(g,f)}_\ell$ is defined as follows:
\begin{equation}\label{eq: right congruence}
  \alpha \equiv^{(g,f)}_\ell \beta \Leftrightarrow S^{(g,f)}_\alpha(\ell)= S^{(g,f)}_\beta(\ell)
\end{equation} for any $\alpha$, $\beta \in \Sast$.
\end{definition}
Clearly, $\equiv^{(g,f)}_\ell$ is a congruence on $\Sast$. In addition, $\equiv^{(g,f)}_\ell$ satisfies
\begin{equation}\label{eq: right congruence}
  \alpha \equiv^{(g,f)}_\ell \beta \Rightarrow \alpha\gamma \equiv^{(g,f)}_\ell \beta\gamma
\end{equation} for any $\alpha$, $\beta$, $\gamma \in \Sast$.\\
\noindent \emph{Proof}: By definition of the congruence $S^{(g,f)}_\alpha(\ell)= S^{(g,f)}_\beta(\ell)$. Then, by (\ref{prop2}), $S_{\alpha\gamma}(\ell)= (S_\gamma \circ S_\alpha)(\ell)= (S_\gamma \circ S_\beta)(\ell)= S_{\beta\gamma}(\ell)$. Thus, $\alpha\gamma \equiv^{(g,f)}_\ell \beta\gamma$ for every $\gamma\in \Sast$. \hfill \qed\\

It concludes that $\equiv^{(g,f)}_\ell$ is a right congruence on $\Sast$. Furthermore, if the quotient set $\Sast / \equiv^{(g,f)}_\ell$ is finite, i.e., $\equiv^{(g,f)}_\ell$ has \emph{finite index}, then the set $Q^{(g,f)}_\ell$ is also finite (and \emph{vice versa}):
\begin{equation}\label{eq: finite index finite states}
  \equiv^{(g,f)}_\ell \mbox{ has finite index }\Leftrightarrow Q^{(g,f)}_\ell=\{S_\alpha(\ell)| \alpha\in \Sast\} \mbox{ is finite}
\end{equation}

\begin{corollary}\label{cor: rigth congruence} Let $\ell\in L$ be an $M$-language. If there exists a factorization on $L$, $(g,f)$, such that the right congruence on $\Sast$ induced by the factorization, $\equiv^{(g,f)}_\ell$, has finite index, then $\ell$ is a recognizable $M$-language.
\end{corollary}
\noindent \emph{Proof}: By (\ref{eq: finite index finite states}), $Q^{(g,f)}_\ell$ is a finite set. Therefore, by Lemma {\ref{theo: recognizability}, $\ell$ is recognized by the $M$-DFA $N^{(g,f)}(\ell, 1)$. \hfill \qed\\

We end this section by providing the main properties of the automaton $N^{(g,f)}(\ell, m)$.

\begin{property}\label{prop: prop of Nerode auto} Let $\ell\in L$ be an $M$-language. Let $(g,f)$ be a factorization on $L$. If the automaton $N^{(g,f)}(\ell, 1)$ is an $M$-DFA then, for any $m \in M$, the $M$-DFA $N^{(g,f)}(\ell, m)$ satisfies the properties:
\begin{enumerate}
  \item $\mathcal{N}^{(g,f)}(\ell, m)= m \cdot \ell$
  \item $N^{(g,f)}(\ell, m)$ is an accessible $M$-DFA.
  \item Each state $S_\alpha(\ell) \in Q^{(g,f)}_\ell$ is a recognizable $M$-language.
  \item For each state $S_\alpha(\ell) \in Q^{(g,f)}_\ell$, $(\mathcal{N}^{(g,f)}(\ell, m))_{S_\alpha(\ell)}= S_\alpha(\ell)$.
  \item $N^{(g,f)}(\ell, m)$ satisfies the necessary condition of minimality ({\bf NcdW}).
\end{enumerate}
\end{property}
\noindent \emph{Proof}:\\
\noindent 1. We recall that $m\in M$ is the initial value in the automaton $N^{(g,f)}(\ell, m)$ that replaces the original initial value $1$ in $N^{(g,f)}(\ell, 1)$. Both automata are exactly the same by excluding those initial values. By condition ({\bf Rcg1}) (see section \ref{sec: M-dfa}), as $N^{(g,f)}(\ell, 1)$ is an $M$-DFA that recognizes $\ell$ (proof of Lemma \ref{theo: recognizability}) then, $N^{(g,f)}(\ell, m)$ is an $M$-DFA that recognizes $m\cdot \ell \in L$ for any $m\in M$.\\

\noindent 2. By Definition \ref{def: nerode automaton} in Lemma \ref{theo: recognizability}, the set $Q^{(g,f)}_\ell=\{S_\alpha(\ell)| \alpha\in \Sast\}$ contains all accessible states from the initial state $S_\varepsilon(\ell)=\ell$ (see the proof below Definition \ref{def: nerode automaton} in Lemma \ref{theo: recognizability}).\\

\noindent 3. As $Q^{(g,f)}_\ell$ is finite then, for any $S_\alpha(\ell)\in Q^{(g,f)}_\ell$, the set $Q^{(g,f)}_{S_\alpha(\ell)}$, that is obtained by replacing $\ell$ by $S_\alpha(\ell)$, is also finite and; thus, each state $S_\alpha(\ell)$ is a recognizable $M$-language by Lemma \ref{theo: recognizability}. In fact, $Q^{(g,f)}_{S_\alpha(\ell)}\subseteq Q^{(g,f)}_\ell$ because, by (\ref{prop2}), for any word $\beta$, $S_{\alpha\beta}(\ell)= (S_\beta \circ S_\alpha)(\ell)= S_\beta(S_\alpha(\ell))\in Q^{(g,f)}_\ell$.\\

\noindent 4. As $S_\alpha(\ell)$ is a state of $N^{(g,f)}(\ell, m)$ then, by (\ref{eq: Mlanguage of Aq}), $(\mathcal{N}^{(g,f)}(\ell, m))_{S_\alpha(\ell)}$ is given by $(\mathcal{N}^{(g,f)}(\ell, m))_{S_\alpha(\ell)}(\beta)= w^\ast(S_\alpha(\ell),\beta)\cdot \rho(S_\alpha(\ell)\beta)$ for any $\beta\in \Sast$, where $w^\ast()$ and $\rho()$ are the extended monoid-transition function and final function provided in (\ref{eq: omega_ast}) and Definition \ref{def: nerode automaton}. Following a similar reasoning as in the proof of Lemma \ref{theo: recognizability}, we have that\\
\noindent $w^\ast(S_\alpha(\ell),\beta)= \prod_{i=1}^{|\beta|} (W_{\beta(i)} \circ S_{\beta[i-1]})(S_\alpha(\ell))= W_\beta(S_\alpha(\ell))= (W_\beta \circ S_\alpha)(\ell)$, and
$\rho(S_\alpha(\ell)\beta)= \rho(S_{\alpha\beta}(\ell))= (S_{\alpha\beta}(\ell))(\varepsilon)= ((S_\beta \circ S_\alpha)(\ell))(\varepsilon)$

\noindent Thus, for any word $\beta$,
\begin{eqnarray*}
\begin{array}{l}
  (\mathcal{N}^{(g,f)}(\ell, m))_{S_\alpha(\ell)}(\beta)= (W_\beta \circ S_\alpha)(\ell)\cdot ((S_\beta \circ S_\alpha)(\ell))(\varepsilon)= \\
  (((W_\beta \circ S_\alpha)\bullet(S_\beta \circ S_\alpha))(\ell))(\varepsilon)= \mbox{ by (\ref{prop5})}\\
  ((\lder{\beta}\circ S_\alpha)(\ell))(\varepsilon)= (\lder{\beta}(S_\alpha(\ell)))(\varepsilon)= S_\alpha(\beta)
\end{array}
\end{eqnarray*}
In conclusion, $(\mathcal{N}^{(g,f)}(\ell, m))_{S_\alpha(\ell)}= S_\alpha(\ell)$.\\

\noindent 5. The necessary condition for minimality ({\bf NcndW}) (see Property \ref{prop: necessary cond min}.3) is fulfilled in the $M$-DFA $N^{(g,f)}(\ell, m)$ because for all states $S_\alpha(\ell)$ and $S_\beta(\ell)$ of that automaton, if $(\mathcal{N}^{(g,f)}(\ell, m))_{S_\alpha(\ell)}= (\mathcal{N}^{(g,f)}(\ell, m))_{S_\beta(\ell)}$ then, by the previous Property \ref{prop: prop of Nerode auto}.4, $S_\alpha(\ell)=S_\beta(\ell)$. \hfill \qed


\section{Natural factorizations on $L(\Sast,M)$}\label{sec: natural factorizations}

Recognizability of $M$-languages provided in section \ref{sec: recognizability} has been based on the hypothesis that, for arbitrary monoids, there exist general factorizations on $L(\Sast,M)$. In this section, we show that each $M$-DFA is able to define a natural factorization on $L$. The construction of a factorization on $L$ induced by an $M$-DFA is based on equalization of transitions (see (\ref{eq: transition-equalized automaton})).

\noindent Let $A=(Q, \Sigma, u, i_u, \delta, w, \rho)$ be an $M$-DFA (Definition \ref{def: M-dfa}). Recall that the $M$-language recognized by $A$ is denoted by $\mathcal{A}$ (see (\ref{eq: Mlanguage of A})), and that $\mathcal{A}_q$ (see \ref{eq: Mlanguage of Aq}) is the $M$-language given for each state $q$ of $A$.\\
\noindent We define the set $P_A=\{((\sigma,q), \lder{\sigma}(\mathcal{A}_q)), ((\varepsilon,u),\mathcal{A})| \sigma\in \Sigma, q\in Q\}$ and the following relation on $P_A$, denoted by $\approx_A$:
\begin{equation}\label{eq: relation equalice}
  ((\sigma,q), \ell)\approx_A ((\tau,p), \ell')\Leftrightarrow \ell=\ell'
\end{equation} for every $((\sigma,q), \ell)$, $((\tau,p), \ell')\in P_A$.\\

By the given definition, $\approx_A$ is an equivalence relation on $P_A$. This equivalence relation has been constructed taken into account the antecedent of the implication given in (\ref{eq: transition-equalized automaton}), and it will allow us to define a factorization on $L$. Clearly, the quotient set $P_A / \approx_A$ contains a finite number of equivalence classes. Let $\pi$ be a function for selecting a unique representative element for each class in $P_A / \approx_A$. Under a given selection $\pi$, each class $C_A \in P_A / \approx_A$, is denoted by $C_A^{\pi}((\sigma,q), \lder{\sigma}(\mathcal{A}_q))$. This notation indicates explicitly the representative element of the class. The given selection $\pi$ is defined is such a way that it always provides the class $C_A^{\pi}((\varepsilon,u),\mathcal{A})$.

\begin{definition}\label{def: factorization induced by A} Let $A=(Q, \Sigma, u, i_u, \delta, w, \rho)$ be an $M$-DFA, let $P_A / \approx_A$ be the quotient set of the equivalence relation $\approx_A$ (\ref{eq: relation equalice}), and let $\pi$ be a selection function of representative elements for the classes in $P_A / \approx_A$. The functions $f_A^{\pi}: L \rightarrow L$ and $g_A^{\pi}: L \rightarrow M$ are defined as follows:\\

\noindent For each $\ell \in L$
\begin{equation}\label{eq: f_A pi}
  \begin{array}{ll}
    f_A^{\pi}(\ell)= \mathcal{A}_u & \mbox{if } \ell= \mathcal{A} ~\wedge~ C_A^{\pi}((\varepsilon,u),\mathcal{A})\in P_A / \approx_A \\
    f_A^{\pi}(\ell)= \mathcal{A}_{q\sigma} & \mbox{if } \ell= \lder{\sigma}(\mathcal{A}_q) ~\wedge~ C_A^{\pi}((\sigma,q), \lder{\sigma}(\mathcal{A}_q)) \in P_A / \approx_A\\
    f_A^{\pi}(\ell)=\ell & \mbox{otherwise}
  \end{array}
\end{equation}
\begin{equation}\label{eq: g_A pi}
  \begin{array}{ll}
    g_A^{\pi}(\ell)= i_u & \mbox{if } \ell= \mathcal{A} ~\wedge~ C_A^{\pi}((\varepsilon,u),\mathcal{A})\in P_A / \approx_A \\
    g_A^{\pi}(\ell)= w(q,\sigma) & \mbox{if } \ell= \lder{\sigma}(\mathcal{A}_q) ~\wedge~ C_A^{\pi}((\sigma,q), \lder{\sigma}(\mathcal{A}_q)) \in P_A / \approx_A\\
    g_A^{\pi}(\ell)= 1 & \mbox{otherwise}
  \end{array}
\end{equation}
\end{definition}

By the given definition, $f_A^\pi$ and $g_A^\pi$ are well defined functions of $F$ and $G$ respectively. We prove that such pair of functions is a factorization on $L$.

\begin{lemma}\label{lem: g_A and f_A factorization} Let $A=(Q, \Sigma, u, i_u, \delta, w, \rho)$ be an $M$-DFA. The pair $(g_A^\pi, f_A^\pi)\in G\times F$ (Definition \ref{def: factorization induced by A}) is a factorization on $L$ induced by the automaton $A$.
\end{lemma}
\noindent \emph{Proof}: For any $\ell\in L$. By Definition \ref{def: factorization induced by A},\\
\begin{itemize}
  \item $\ell=\mathcal{A}$. Then, $g_A^\pi(\ell)\cdot f_A^\pi(\ell)= i_u\cdot \mathcal{A}_u= \mathcal{A}$ (by (\ref{eq: Mlanguage of Aq}) and (\ref{eq: Mlanguage of A})).
  \item $\ell=\lder{\tau}(\mathcal{A}_p)$ for some $\tau\in \Sigma$ and $p\in Q$, then there exists some class $C_A^\pi \in P_A/\approx_A$ such that $\ell= \lder{\sigma}(\mathcal{A}_q)$ or $\ell=\mathcal{A}$. This last case is the same as the given above. For the former case, assume that $((\sigma,q),\lder{\sigma}(\mathcal{A}_q))$ is the given representative by $\pi$. Then, $g_A^\pi(\ell)\cdot f_A^\pi(\ell)= w(q,\sigma)\cdot \mathcal{A}_{q\sigma}= \lder{\sigma}(\mathcal{A}_q)=\ell$ (by (\ref{eq: Mlanguage derivative alpha Aq})).
  \item Otherwise, $g_A^\pi(\ell)\cdot f_A^\pi(\ell)=1 \cdot \ell$.
\end{itemize}
Therefore, $g_A^\pi \bullet f_A^\pi= f_e$. By Definition \ref{def: factorization}, the pair $(g_A^\pi, f_A^\pi)\in G\times F$ is a factorization on $L$. \hfill \qed\\

Let us observe that the pair $(g_A^\pi, f_A^\pi)$ depends on the selection function $\pi$, and that Lemma \ref{lem: g_A and f_A factorization} holds for any selection $\pi$. Given an $M$-DFA $A=(Q, \Sigma, u, i_u, \delta, w, \rho)$, define the sets $\widehat{P}_A=\{\mathcal{A},\lder{\sigma}(\mathcal{A}_q)| \sigma\in \Sigma, q\in Q\}$ and $\widehat{Q}_A=\{\mathcal{A}_q| q\in Q\}$. Any factorization on $L$ induced by $A$, $(g_A^\pi, f_A^\pi)$, satisfies that
\begin{equation}\label{eq: image of gpi fpi}
\begin{array}{ll}
f_A^\pi(\ell)\in \widehat{Q}_A & \mbox{if } \ell \in \widehat{P}_A\\
f_A^\pi(\ell)= \ell & \mbox{otherwise }
\end{array}
\end{equation} for any $\ell \in L$.

\begin{lemma}\label{lem: finite index by f_A} Let $A=(Q, \Sigma, u, i_u, \delta, w, \rho)$ be an $M$-DFA. For any factorization on $L$ induced by $A$, $(g_A^\pi, f_A^\pi)$, the right congruence $\equiv_{f_A^\pi(\mathcal{A})}^{(g_A^\pi, f_A^\pi)}$ has finite index.
\end{lemma}
\noindent \emph{Proof}: Recall that $\mathcal{A}$ is the $M$-language recognized by $A$. As $A$ is an $M$-DFA, the set of states $Q$ is finite, and the sets of $M$-languages $\widehat{Q}_A=\{\mathcal{A}_q| q\in Q\}$  and $\widehat{P}_A=\{\mathcal{A},\lder{\sigma}(\mathcal{A}_q)| \sigma\in \Sigma, q\in Q\}$ are also finite. 
We prove that for any $\alpha \in \Sast$,
\begin{equation}\label{eq: S_alpha in hat(Q)}
S^{(g_A^\pi, f_A^\pi)}_\alpha(f_A^\pi(\mathcal{A}))\in \widehat{Q}_A
\end{equation}

For convenience, we omit the superscript $(g_A^\pi, f_A^\pi)$ in $S^{(g_A^\pi, f_A^\pi)}_\alpha$.
\noindent By induction on the prefixes of $\alpha$:\\

\noindent - Basis. By (\ref{prop1}), $S_\varepsilon(f_A^\pi(\mathcal{A}))= f_A^\pi(\mathcal{A})$. As $\mathcal{A}\in \widehat{P}_A$, then, by (\ref{eq: image of gpi fpi}), $S_\varepsilon(f_A^\pi(\mathcal{A}))\in \widehat{Q}_A$.\\
\noindent - Hypothesis. Let us assume that $S_{\alpha'}(f_A^\pi(\mathcal{A}))\in \widehat{Q}_A$ where $\alpha'$ is a prefix of $\alpha$.\\
\noindent - Induction step. Let us consider $\alpha'\sigma$ be a prefix of $\alpha$ with $\sigma \in \Sigma$. By (\ref{prop2}) and (\ref{prop1b}), $S_{\alpha'\sigma}(f_A^\pi(\mathcal{A}))=(S_\sigma \circ S_{\alpha'})(f_A^\pi(\mathcal{A}))= f_A^\pi(\lder{\sigma}(S_{\alpha'}(f_A^\pi(\mathcal{A}))))$. By induction hypothesis, $S_{\alpha'}(f_A^\pi(\mathcal{A}))\in \widehat{Q}_A$, i.e., there exists some $q\in Q$ such that $S_{\alpha'}(f_A^\pi(\mathcal{A}))=\mathcal{A}_q$. By substitution, $S_{\alpha'\sigma}(f_A^\pi(\mathcal{A}))= f_A^\pi(\lder{\sigma}(\mathcal{A}_q))$. As $\lder{\sigma}(\mathcal{A}_q)\in \widehat{P}_A$, then, by (\ref{eq: image of gpi fpi}), $f_A^\pi(\lder{\sigma}(\mathcal{A}_q))\in \widehat{Q}_A$. Therefore, $S_{\alpha'\sigma}(f_A^\pi(\mathcal{A}))\in \widehat{Q}_A$.\\

\noindent This fact proves that the set $Q_{f_A^\pi(\mathcal{A})}^{(g_A^\pi, f_A^\pi)}=\{S_\alpha(f_A^\pi(\mathcal{A}))| \alpha\in \Sast\}$ is finite because $Q_{f_A^\pi(\mathcal{A})}^{(g_A^\pi, f_A^\pi)}\subseteq \widehat{Q}_A$. Therefore, by (\ref{eq: finite index finite states}), $\equiv_{f_A^\pi(\mathcal{A})}^{(g_A^\pi, f_A^\pi)}$ has finite index. \hfill \qed

\begin{remark}\label{rem: equivalence A and Nerode f(A)} Given an $M$-DFA $A=(Q, \Sigma, u, i_u, \delta, w, \rho)$, Lemma \ref{lem: finite index by f_A} states that $\equiv_{f_A^\pi(\mathcal{A})}^{(g_A^\pi, f_A^\pi)}$ has finite index. By Corollary \ref{cor: rigth congruence} and Lemma \ref{theo: recognizability}, the $M$-DFA $N^{(g_A^\pi, f_A^\pi)}(f_A^\pi(\mathcal{A}),1)$ recognizes the $M$-language $f_A^\pi(\mathcal{A})$.\\
By Property \ref{prop: prop of Nerode auto}.1, $N^{(g_A^\pi, f_A^\pi)}(f_A^\pi(\mathcal{A}),g_A^\pi(\mathcal{A}))$ recognizes $g_A^\pi(\mathcal{A})\cdot f_A^\pi(\mathcal{A})$. By Definition \ref{def: factorization induced by A} (or Lemma \ref{lem: g_A and f_A factorization}), $g_A^\pi(\mathcal{A})\cdot f_A^\pi(\mathcal{A})= i_u\cdot \mathcal{A}_u= \mathcal{A}$.\\
Therefore, $N^{(g_A^\pi, f_A^\pi)}(f_A^\pi(\mathcal{A}),g_A^\pi(\mathcal{A}))$ is equivalent to $A$.
\end{remark}

Previous Remark shows that $N^{(g_A^\pi, f_A^\pi)}(f_A^\pi(\mathcal{A}),g_A^\pi(\mathcal{A}))$ is equivalent to $A$. Those automata could be very different because the former one satisfies all the properties given in Property \ref{prop: prop of Nerode auto} but, however, these properties may be absent in the automaton $A$. The $M$-DFA $N^{(g_A^\pi, f_A^\pi)}(f_A^\pi(\mathcal{A}),g_A^\pi(\mathcal{A}))$ has their transitions equalized. This is derived by the way factorization $(g_A^\pi, f_A^\pi)$ is constructed. The next property collects the main properties involved any factorization on $L$ induced by $A$.

\begin{property}\label{prop: prop of natural fact}
 Let $A=(Q, \Sigma, u, i_u, \delta, w, \rho)$ be an $M$-DFA. For any factorization on $L$ induced by $A$, $(g_A^\pi, f_A^\pi)$, the following properties hold:
\begin{enumerate}
\item $N^{(g_A^\pi, f_A^\pi)}(f_A^\pi(\mathcal{A}),g_A^\pi(\mathcal{A}))$ is equivalent to $A$.
\item $N^{(g_A^\pi, f_A^\pi)}(f_A^\pi(\mathcal{A}),g_A^\pi(\mathcal{A}))$ is transition-equalized.
\item If $A$ is minimal then $N^{(g_A^\pi, f_A^\pi)}(f_A^\pi(\mathcal{A}),g_A^\pi(\mathcal{A}))$ is minimal.
\item There is a minimal and transition-equalized $M$-DFA equivalent to $A$.
\item If $A$ is minimal then $f_A^\pi$ is idempotent, i.e., $f_A^\pi \circ f_A^\pi= f_A^\pi$
\item If $A$ is minimal and transition-equalized then $f_A^{\pi_i}=f_A^{\pi_j}$ for any selection functions $\pi_i$ and $\pi_j$.
\end{enumerate}
\end{property}
\noindent \emph{Proof}:\\
\noindent 1. The equivalence is shown in Remark \ref{rem: equivalence A and Nerode f(A)}.\\

\noindent 2. Let $N^{(g_A^\pi, f_A^\pi)}(f_A^\pi(\mathcal{A}),g_A^\pi(\mathcal{A}))= (Q^{(g_A^\pi, f_A^\pi)}_{f_A^\pi(\mathcal{A})}, \Sigma, f_A^\pi(\mathcal{A}),g_A^\pi(\mathcal{A}),\delta, w, \rho )$ be the structure of such automaton provided in Definition \ref{def: nerode automaton}. Let us consider two states $S_\alpha(f_A^\pi(\mathcal{A}))$ and $S_\beta(f_A^\pi(\mathcal{A}))$ and two symbols $\sigma$, $\tau \in \Sigma$. By (\ref{eq: S_alpha in hat(Q)}), $S_\alpha(f_A^\pi(\mathcal{A}))$, $S_\beta(f_A^\pi(\mathcal{A}))\in \widehat{Q}_A$, i.e., there are two states $q$, $p\in Q$, such that $S_\alpha(f_A^\pi(\mathcal{A}))=\mathcal{A}_q$ and $S_\beta(f_A^\pi(\mathcal{A}))= \mathcal{A}_p$. Now consider (\ref{eq: transition-equalized automaton}):
\begin{eqnarray*}
\begin{array}{l}
  \lder{\sigma}(\mathcal{N}^{(g_A^\pi, f_A^\pi)}(f_A^\pi(\mathcal{A}),g_A^\pi(\mathcal{A}))_{S_\alpha(f_A^\pi(\mathcal{A}))})= \lder{\tau}(\mathcal{N}^{(g_A^\pi, f_A^\pi)}(f_A^\pi(\mathcal{A}),g_A^\pi(\mathcal{A}))_{S_\beta(f_A^\pi(\mathcal{A}))}),\\
  \mbox {by Property \ref{prop: prop of Nerode auto}.4}\\
  \lder{\sigma}(S_\alpha(f_A^\pi(\mathcal{A})))= \lder{\tau}(S_\beta(f_A^\pi(\mathcal{A})))\\
  \mbox{then, }\\
  \lder{\sigma}(\mathcal{A}_q)= \lder{\tau}(\mathcal{A}_p)
\end{array}
\end{eqnarray*}

\noindent This fact implies that $((\sigma,q),\lder{\sigma}(\mathcal{A}_q))$ and $((\tau,p),\lder{\tau}(\mathcal{A}_p))\in C^{\pi}_A$, i.e, they are in the same class of $P_A/\approx_A$. Then, by Definition \ref{def: factorization induced by A}, $f^{\pi}_A(\lder{\sigma}(\mathcal{A}_q))= f^{\pi}_A(\lder{\tau}(\mathcal{A}_p))$ and $g^{\pi}_A(\lder{\sigma}(\mathcal{A}_q))=g^{\pi}_A(\lder{\tau}(\mathcal{A}_p))$.\\

\noindent Thus, $\delta(S_\alpha(f_A^\pi(\mathcal{A})),\sigma)= S_{\alpha\sigma}(f_A^\pi(\mathcal{A}))=(S_\sigma \circ S_\alpha)(f_A^\pi(\mathcal{A}))= f^{\pi}_A(\lder{\sigma}(\mathcal{A}_q))= f^{\pi}_A(\lder{\tau}(\mathcal{A}_p))= \delta(S_\beta(f_A^\pi(\mathcal{A})),\tau)$. In a similar way,
$w(S_\alpha(f_A^\pi(\mathcal{A})),\sigma)= (W_\sigma \circ S_\alpha)(f_A^\pi(\mathcal{A}))= g^{\pi}_A(\lder{\sigma}(\mathcal{A}_q))= g^{\pi}_A(\lder{\tau}(\mathcal{A}_p))= w(S_\beta(f_A^\pi(\mathcal{A})),\tau)$. In conclusion,
$N^{(g_A^\pi, f_A^\pi)}(f_A^\pi(\mathcal{A}),g_A^\pi(\mathcal{A}))$ is transition-equalized.\\

\noindent 3. By the proof in Lemma \ref{lem: finite index by f_A}, $Q^{(g_A^\pi, f_A^\pi)}_{f_A^\pi(\mathcal{A})}\subseteq \widehat{Q}_A$. By definition $\|\widehat{Q}_A\|\leq \|Q\|$ where $Q$ is the set of states of the minimal $M$-DFA $A$. Then, $\|Q^{(g_A^\pi, f_A^\pi)}_{f_A^\pi(\mathcal{A})}\| \leq \|Q\|$. Therefore, $N^{(g_A^\pi, f_A^\pi)}(f_A^\pi(\mathcal{A}),g_A^\pi(\mathcal{A}))$ is a minimal $M$-DFA.\\

\noindent 4. It is a consequence of the previous properties.\\

\noindent 5. This property is proved by exhaustive case analysis and the application of the condition {\bf NcndS} (Property \ref{prop: necessary cond min}.2) for a minimal $M$-DFA. We omit the proof by brevity.\\

\noindent 6. By Definition \ref{def: factorization induced by A} and (\ref{eq: image of gpi fpi}):
\begin{quote}
- If $\ell \notin \widehat{P}_A$, then $f^{\pi_i}_A(\ell)=f^{\pi_j}_A(\ell)=\ell$.\\
- If $\ell \in \widehat{P}_A$ and $\ell=\mathcal{A}$, then $f^{\pi_i}_A(\ell)=f^{\pi_j}_A(\ell)=\mathcal{A}_u$.\\
- If $\ell \in \widehat{P}_A$ and $\ell= \lder{\sigma}(A_q)$ and $C^{\pi_i}_A((\sigma,q),\lder{\sigma}(A_q))\in P_A/\approx_A$, then $C^{\pi_i}_A((\sigma,q),\lder{\sigma}(A_q))= C^{\pi_j}_A((\tau,p),\lder{\tau}(A_p))$ with $\lder{\sigma}(A_q)= \lder{\tau}(A_p)$. As $A$ is transition-equalized, by (\ref{eq: transition-equalized automaton}), $q\sigma=p\tau$. Therefore, $f^{\pi_i}_A(\ell)=\mathcal{A}_{q\sigma}= \mathcal{A}_{p\tau}=f^{\pi_j}_A(\ell)$.
\end{quote} \hfill \qed

\section{General Myhill-Nerode Theorem}\label{sec: main result}

Results presented in previous sections allow us to enunciate a general theorem for recognizability of $M$-languages.

\begin{theorem}\label{theo: main result}({\bf General Myhill-Nerode Theorem}) Let $\ell \in L$ be an $M$-language. The following two conditions are equivalent:
\begin{quote}
(i) $\ell$ is a recognizable $M$-language;\\
(ii) there exists a factorization on $L$, $(g,f)$, such that the right congruence $\equiv_{f(\ell)}^{(g,f)}$ has finite index.
\end{quote}
\end{theorem}
\noindent \emph{Proof}:\\
\noindent (i) $\Rightarrow$ (ii). If $\ell$ is a recognizable $M$-language then there is an $M$-DFA $A=(Q, \Sigma, u, i_u, \delta, w, \rho)$ such that $\mathcal{A}=\ell$. By Lemma \ref{lem: g_A and f_A factorization}, the pair $(g_A^\pi, f_A^\pi)\in G\times F$ (Definition \ref{def: factorization induced by A}) is a factorization on $L$ induced by the automaton $A$. By Lemma \ref{lem: finite index by f_A}, the right congruence $\equiv_{f_A^\pi(\mathcal{A})}^{(g_A^\pi, f_A^\pi)}$ has finite index. In addition, by Remark \ref{rem: equivalence A and Nerode f(A)}, the $M$-DFA $N^{(g_A^\pi, f_A^\pi)}(f_A^\pi(\mathcal{A}),g_A^\pi(\mathcal{A}))$ recognizes $\mathcal{A}=\ell$.\\

\noindent (ii) $\Rightarrow$ (i). If there exists a factorization on $L$, $(g,f)$, such that the right congruence $\equiv_{f(\ell)}^{(g,f)}$ has finite index then, by Lemma \ref{theo: recognizability}, $f(\ell)$ is recognized by the $M$-DFA $N^{(g,f)}(f(\ell), 1)$. By Property \ref{prop: prop of Nerode auto}.1, the $M$-DFA $N^{(g,f)}(f(\ell), g(\ell))$ recognizes the $M$-language $g(\ell)\cdot f(\ell)= \ell$ since $(g,f)$ is a factorization on $L$. \hfill \qed

\section{Recognition Capability of factorizations on $L(\Sast,M)$}\label{sec: recogn capability}
For a monoid $M$ and an alphabet $\Sigma$, the set of all recognizable $M$-languages in $L(\Sast,M)$ is denoted by $\mathcal{R}_L$.
By following the previous general Myhill-Nerode Theorem, we can define the \emph{recognition capability} of a factorization on $L$, $(g,f)$, as the set
\begin{equation}\label{eq: RecogCap}
  \mbox{RcgCap}^{(g,f)}_L=\{\ell \in L~|~ N^{(g,f)}(f(\ell),1)\mbox{ is an } M\mbox{-DFA}\}
\end{equation}
This set, $\mbox{RcgCap}^{(g,f)}_L$, represents the set of all recognizable $M$-languages for a given factorization on $L$, i.e, the languages recognized by $N^{(g,f)}(f(\ell),m)$ for some $m\in M$. Obviously, $\mbox{RcgCap}^{(g,f)}_L \subseteq \mathcal{R}_L$. In this section, we study the recognition capability of three types of factorizations: (a) trivial factorization; (b) maximal factorizations; and (c) composition of natural factorizations.

\subsection{Case of study: Trivial factorization} For every monoid $M$, $L(\Sast, M)$ admits the trivial factorization $(g_e,f_e)$, which is the identity element of $G\times F$ under composition of factorizations. By (\ref{eq: def of S(g,f)alpha}) and (\ref{eq: def of W(g,f)alpha}), for any $\ell\in L$ and $\alpha\in \Sast$, $S^{(g_e,f_e)}_\alpha(\ell)=\lder{\alpha}(\ell)$ and $W^{(g_e,f_e)}_\alpha(\ell)=1$. In addition, $f_e(\ell)=\ell$.\\
Let us consider that $N^{(g_e,f_e)}(\ell,1)$ is an $M$-DFA for a language $\ell\in L$. By Definition \ref{def: nerode automaton}, $Q^{(g_e,f_e)}_\ell=\{\lder{\alpha}(\ell)| \alpha \in \Sast\}$ is finite and its monoid-transition function satisfies that $w(\lder{\alpha},\sigma)=1$ for any state and symbol.\\
The $M$-DFA $N^{(g_e,f_e)}(\ell,1)$, which recognizes $\ell$, is merely and ordinary DFA equipped with a final-state function $\rho$. Furthermore, by (\ref{eq: Mlanguage of A}), $(\mathcal{N}^{(g_e,f_e)}(\ell,1))(\beta)= \rho(\lder{\beta}(\ell))= (\lder{\beta}(\ell))(\varepsilon)=\ell(\beta)$. Therefore, as $Q^{(g_e,f_e)}_\ell$ is finite, $\ell$ has finite rank. In conclusion, only $M$-languages of finite rank may be recognized by using the trivial factorization.\\
Previous discussion allows us to introduce the notion of a $\rho$-DFA, i.e., an ordinary DFA with a final-state function $\rho$.
We can represent a $\rho$-DFA by $A=(Q, \Sigma, u, \delta, \rho)$ where $(Q, \Sigma, u, \delta)$ is a $DFA$ and $\rho: Q\rightarrow M$. A $\rho$-DFA $A$, recognizes the $M$-language $\mathcal{A}(\alpha)=\rho(u\alpha)$ for any word $\alpha$. One can define the notion of minimal $\rho$-DFA, and by Property \ref{prop: necessary cond min}.3 and Property \ref{prop: suficcient cond min}, it is concluded that the sufficient and necessary condition for an accessible $\rho$-DFA to be minimal is just the condition {\bf NcndW}, i.e., $A$ is a minimal $\rho$-DFA if and only if for every $p$, $q\in Q$, $\mathcal{A}_p=\mathcal{A}_q \Rightarrow p=q$. Therefore, $N^{(g_e,f_e)}(\ell,1)$, when it is finite, is the minimal $\rho$-DFA that recognizes $\ell$ since $N^{(g_e,f_e)}(\ell,1)$ satisfies Property \ref{prop: necessary cond min}.3 ({\bf NcdW}). However, this fact does not prevent to find another factorization on $L$ that provides an $M$-DFA with lesser states than a minimal $\rho$-DFA (see fig.2 and fig. 3 in \cite{mendivil2018}).\\
Let us observe that $N^{(g_e,f_e)}(\ell,m)= N^{(g_e,f_e)}(m\cdot \ell,1)$ since $\ell$ is of finite rank and $N^{(g_e,f_e)}(\ell,m)$ is a $\rho$-DFA. Therefore, for any monoid $M$,\\
$\mathcal{R}_L \supseteq \mbox{RcgCap}^{(g_e,f_e)}_L=\{\ell\in L | \ell \mbox{ is recognized by a } \rho\mbox{-DFA}\}$\\

In the context of fuzzy automata, $\rho$-DFAs are called crisp deterministic fuzzy automata. This kind of automata has been studied by Ignatovi\'c et al. \cite{Jelena2010} from the perspective of the Myhill-Nerode Theorem and minimization algorithms for crisp deterministic fuzzy automata.

\subsection{Case of study: Maximal factorizations} Let us consider that $L(\Sast,M)$ admits a maximal factorization. By the results in \cite{Gerdjikov2019}, if $M$ contains a \emph{zero} element then $(M,\cdot, 1)$ is \emph{zero-divisor-free}. In addition, $(M\setminus\{0\},\cdot,1)$ is also a monoid. In particular, \cite{Gerdjikov2019} studies \emph{mge}-monoids\footnote{An \emph{mge}-monoid satisfies left and right cancellation axioms and the right most general equalizer axiom \cite{Gerdjikov2019}.} and their conditions to obtain maximal factorizations. In order to simplify this case of study, we consider that $(M,\cdot, 1)$ is a monoid without a zero element.\\
A maximal factorization on $L$, $(g_h,h)\in G\times F$, is a factorization that satisfies $h(m\cdot \ell)=h(\ell)$ for any $\ell \in L$ and $m \in M$. This strong property implies that $h$ is idempotent: $h(\ell)=h(g_h(\ell)\cdot h(\ell))= h(h(\ell))$ for any $\ell\in L$.\\
Let us consider that $N^{(g_h,h)}(h(\ell),1)$ is an $M$-DFA for a language $\ell\in L$.\\
By (\ref{eq: def of S(g,f)alpha}) for any $\ell\in L$ and $\alpha\in \Sast$, $S^{(g_h,h)}_\alpha(h(\ell))=h(\lder{\alpha}(h(\ell)))$. The reader may prove the given expression by induction on $|\alpha|$ by using that $h$ is idempotent and a maximal factorization. The hint is to apply equation (\ref{eq: basic prop fact1}) when proving the induction step. This expression simplifies the definition of $W^{(g_h,h)}_\alpha(h(\ell))$ in (\ref{eq: def of W(g,f)alpha}). Then, the set of states of $N^{(g_h,h)}(h(\ell),1)$ is the set $Q^{(g_h,h)}_{h(\ell)}=\{h(\lder{\alpha}(h(\ell))) | \alpha\in \Sast\}$.\\
Let us recall that by Property \ref{prop: prop of Nerode auto}.1, $N^{(g_h,h)}(h(\ell),g(\ell))$ recognizes $\ell\in L$. The interesting aspect of this automata is that it satisfies the sufficient condition of minimality (Property \ref{prop: suficcient cond min}).
\begin{property}\label{prop: mimimal by maxfact} Let $(g_h,h)$ be a maximal factorization on $L$. If $N^{(g_h,h)}(h(\ell),g(\ell))$ is an $M$-DFA then it is a minimal $M$-DFA that recognizes $\ell\in L$.
\end{property}
\noindent \emph{Proof}: Let us consider an $M$-DFA $A=(Q,\Sigma, u, i_u, \delta, w, \rho)$ such that it is accessible, but it does not satisfy the sufficient condition for minimality (see Property \ref{prop: suficcient cond min}): for two words $\alpha$, $\beta\in \Sast$, two values $m$, $m'\in M$, and an $M$-language $\ell' \in L$, $u\alpha\neq u\beta \wedge \lder{\alpha}(\mathcal{A})= m\cdot \ell' \wedge \lder{\beta}(\mathcal{A})= m'\cdot \ell'$.\\
\noindent As $(g_h,h)$ is a maximal factorization, $h(\lder{\alpha}(\mathcal{A}))=h(\lder{\beta}(\mathcal{A}))$ because $h(m\cdot \ell')= h(m' \cdot \ell')= h(\ell')$. By (\ref{eq: Mlanguage derivative alpha Aq}), $\lder{\alpha}(\mathcal{A})= i_u \cdot w^\ast(u,\alpha) \cdot \mathcal{A}_{u\alpha}$. Then, $h(\lder{\alpha}(\mathcal{A}))= h(\mathcal{A}_{u\alpha})$. Similarly, $h(\lder{\beta}(\mathcal{A}))= h(\mathcal{A}_{u\beta})$. Therefore, $h(\mathcal{A}_{u\alpha})= h(\mathcal{A}_{u\beta})$. By identifying, $A$ with $N^{(g_h,h)}(h(\ell),g(\ell))$, which is accessible, then $u\alpha= S_\alpha(h(\ell))= h(\lder{\alpha}(h(\ell)))$ and $u\beta= h(\lder{\beta}(h(\ell)))$. By Property \ref{prop: prop of Nerode auto}.4, as $\mathcal{A}_{u\alpha}=u\alpha$ and $\mathcal{A}_{u\beta}=u\beta$, then $h(h(\lder{\alpha}(h(\ell))))=h(h(\lder{\beta}(h(\ell))))$. As $h()$ is idempotent, then $u\alpha=u\beta$ following our identification. This is a contradiction with the initial hypothesis. Therefore, $N^{(g_h,h)}(h(\ell),g(\ell))$ satisfies the sufficient condition for minimality. \hfill \qed\\

A maximal factorization on $L(\Sast, M)$ achieves the maximal recognition capability.
\begin{lemma}\label{lem: recgacp by maxfact} Let $(g_h,h)$ be a maximal factorization on $L$. Then, $\mathcal{R}_L = \mbox{RcgCap}^{(g_h,h)}_L$
\end{lemma}
\noindent \emph{Proof}: Let $\ell \in \mathcal{R}_L$ be recognized by an $M$-DFA $A=(Q,\Sigma, u, i_u, \delta, w, \rho)$, i.e., $\mathcal{A}=\ell$. Without loss of generality, $A$ is accessible, i.e, $Q=\{u\alpha~| \alpha\in \Sast\}$. The set $\widehat{Q}_A=\{\mathcal{A}_{u\alpha}~| \alpha\in\Sast\}$ is finite. Thus, $h(\widehat{Q}_A)=\{h(\mathcal{A}_{u\alpha})~| \alpha\in\Sast\}$ is also finite. By (\ref{eq: Mlanguage derivative alpha Aq}), $\lder{\alpha}(\mathcal{A})= i_u \cdot w^\ast(u,\alpha) \cdot \mathcal{A}_{u\alpha}$. Then, as $(g_h,h)$ is a maximal factorization, $h(\lder{\alpha}(\mathcal{A}))=h(\mathcal{A}_{u\alpha})$. As $\mathcal{A}= g_h(\mathcal{A})\cdot h(\mathcal{A})$; by (\ref{eq: basic prop fact1}),\\
\noindent $h(\lder{\alpha}(\mathcal{A}))= h(\lder{\alpha}(g_h(\mathcal{A})\cdot h(\mathcal{A})))= h(g_h(\mathcal{A})\cdot \lder{\alpha}(h(\mathcal{A})))= h(\lder{\alpha}(h(\mathcal{A})))$. Therefore, the set $\{h(\lder{\alpha}(h(\ell)))~| \alpha\in \Sast\}$ is finite. This set is the set of states of the $M$-DFA $N^{(g_h,h)}(h(\ell),g(\ell))$ which recognizes $\ell$. By the  definitions given at the beginning of this section, $\ell \in \mbox{RcgCap}^{(g_h,h)}_L$, i.e., $\mathcal{R}_L \subseteq  \mbox{RcgCap}^{(g_h,h)}_L$; thus, $\mathcal{R}_L = \mbox{RcgCap}^{(g_h,h)}_L$. \hfill \qed\\

When the maximal factorization has an explicit formulae, it is possible to construct determinization and minimization algorithms for automata. Maximal factorizations produce very efficient constructions. Kirsten and M\"aurer \cite{Kirsten2005} show that their determinization algorithm of weighted automata is optimal using maximal factorizations and the zero-divisor-free condition (Theorem 3.3 in \cite{Kirsten2005}). This behaviour has been corroborated in some determinization methods for fuzzy automata \cite{Mendivil2014b}\cite{Stanimirovic2017}. The original Mohri's minimization algorithm for weighted automata over tropical semiring applies a maximal factorization \cite{Mohri1997}. Other examples of applications of maximal factorizations are in \cite{Eisner2003}\cite{mendivil2018}\cite{Stanimirovic2017b}\cite{mendivil2016}.

\subsection{Case of study: Composition of Natural factorizations} Let us consider an arbitrary monoid $M$. Let $A$ be an $M$-DFA. By Property \ref{prop: prop of natural fact}.4, there exists a minimal and transition-equalized $M$-DFA equivalent to $A$. Thus, we consider that $A$ is a minimal and transition-equalized $M$-DFA. By Property \ref{prop: prop of natural fact}.6, the factorization on $L$ induced by $A$ is unique. That factorization is simply denoted $(g_A,f_A)$. By Lemma \ref{lem: finite index by f_A} and Remark \ref{rem: equivalence A and Nerode f(A)}, $\mbox{RcgCap}^{(g_A,f_A)}_L$ is not empty since $f_A(\mathcal{A})$ is in this set. Let us recall that, by Lemma \ref{lem: composition of factorizations}, the composition of factorizations on $L$ is again a factorization on $L$. We study the composition of natural factorizations to provide the result that the composition preserves the recognition capability of each individual natural factorization.

\begin{lemma}\label{lemma: composition of natural fact}
Let $\{Ak\}_{k=1..n}$ be a finite family of $n$ minimal and transition-equalized $M$-DFAs; and, let $\{(g_{Ak},f_{Ak})\}_{k=1..n}$ be the famility of the natural factorizations on $L$ induced by those automata. The factorization on $L$, $(g,f)=[(g_{Ak},f_{Ak})]_1^n$, obtained by the composition of the family $\{(g_{Ak},f_{Ak})\}_{k=1..n}$, satisfies the next property,
\begin{equation}\label{eq: lemma composition nat fact}
  \mbox{RcgCap}^{(g,f)}_L \supseteq \bigcup_{k=1}^n \mbox{RcgCap}^{(g_{Ak},f_{Ak})}_L
\end{equation}
\end{lemma}
\noindent \emph{Proof}: For each $M$-DFA $Ak=(Q_{Ak}, \Sigma, uk,i_{uk}, \delta_k, w_k, \rho_k)$, with $k:1..n$: $Q_{Ak}$ is the set of states; $\widehat{Q}_{Ak}=\{\mathcal{A}k_{q}~| q\in Q_{Ak}\}$; and $\widehat{P}_{Ak}=\{\mathcal{A}k, \lder{\sigma}(\mathcal{A}k_q)~| q\in Q_{Ak}, \sigma \in \Sigma\}$. By definition of composition of factorizations on $L$, $f= f_{An}\circ ...\circ f_{A1}$. By (\ref{eq: image of gpi fpi}), it is simple to show that,
\begin{equation}\label{eq: general image of fAk}
\begin{array}{ll}
f(\ell)\in \bigcup_{k=1}^n \widehat{Q}_{Ak} & \mbox{if } \ell \in \bigcup_{k=1}^n \widehat{P}_{Ak}\\
f(\ell)= \ell & \mbox{otherwise }
\end{array}
\end{equation} for any $\ell \in L$. Let us observe that if $\ell \notin \bigcup_{k=1}^n \widehat{P}_{Ak}$ then $f(\ell)=\ell$. Let $j$, $n\geq j\geq 1$, be the first index such that $\ell \in \widehat{P}_{Aj}$, then $f(\ell)\in \bigcup_{k=j}^n \widehat{Q}_{Ak}$. This is so because, by (\ref{eq: image of gpi fpi}), $(f_{Aj-1}\circ...\circ f_{A1})(\ell)=\ell$, $f_{Aj}(\ell) \in \widehat{Q}_{Aj}$ and $f_{Aj}(\ell)$ may belong (or not) to any $\widehat{P}_{Ak}$ with $k:j+1..n$.\\

\noindent Let us consider an arbitrary $\ell\in L$ such that, for some arbitrary $Aj$, with $1\leq j\leq n$, $\ell \in \mbox{RcgCap}^{(g_{Aj},f_{Aj})}_L$. That is, the $M$-DFA $N^{(g_{Aj},f_{Aj})}(f_{Aj}(\ell),1)$ recognizes $f_{Aj}(\ell)$. The finite set of states of this automaton is $Q^{(g_{Aj},f_{Aj})}_{f_{Aj}(\ell)}=\{S^{(g_{Aj},f_{Aj})}_\alpha(f_{Aj}(\ell))~| \alpha\in \Sast \}$. By (\ref{eq: def of S(g,f)alpha}) and (\ref{eq: image of gpi fpi}), it is simple to show that $S^{(g_{Aj},f_{Aj})}_\alpha(f_{Aj}(\ell))\in \widehat{Q}_{Aj}$ or $S^{(g_{Aj},f_{Aj})}_\alpha(f_{Aj}(\ell))=\lder{\alpha}(\ell)$. This last case happens when each language in the composition of $S^{(g_{Aj},f_{Aj})}_\alpha(f_{Aj}(\ell))$ does not belong to $\widehat{P}_{Aj}$ and $f_{Aj}$ behaves like the identity $f_e$. This structure of $Q^{(g_{Aj},f_{Aj})}_{f_{Aj}(\ell)}$ is important in the next step of the proof.\\
\noindent We claim that the following property holds for the composition and the language $\ell$,
\begin{equation}\label{eq: claim}
  S^{(g,f)}_\alpha(f(\ell)) \in  (\bigcup_{k=1}^{n} \widehat{Q}_{Ak}) \cup Q^{(g_{Aj},f_{Aj})}_{f_{Aj}(\ell)}
\end{equation} for any $\alpha \in \Sast$.\\

\noindent By induction on the length of $\alpha \in \Sast$:\\

\noindent - Basis. Let $\alpha=\varepsilon$. By (\ref{prop1}), $S^{(g,f)}_\varepsilon(f(\ell))=f(\ell)$. By (\ref{eq: general image of fAk}), $f(\ell)\in \bigcup_{k=1}^n \widehat{Q}_{Ak}$ or $f(\ell)=\ell$. In this last case, $\ell \notin \widehat{P}_{Aj}$. This implies that, by (\ref{prop1}) and (\ref{eq: image of gpi fpi}), $S^{(g_{Aj},f_{Aj})}_\varepsilon(f_{Aj}(\ell))= S^{(g_{Aj},f_{Aj})}_\varepsilon(\ell)=\ell \in Q^{(g_{Aj},f_{Aj})}_{f_{Aj}(\ell)}$. The property holds.\\
\noindent - Hypothesis. Let us assume that (\ref{eq: claim}) is valid for $\alpha \in \Sast$.\\
\noindent - Induction Step. Let $\alpha'=\alpha\sigma$ with $\alpha\in \Sast$ and $\sigma\in \Sigma$. By (\ref{prop1b}) and (\ref{prop2}), $S^{(g,f)}_{\alpha\sigma}(f(\ell))= (S^{(g,f)}_\sigma \circ S^{(g,f)}_\alpha)(f(\ell))= f(\lder{\sigma}(S^{(g,f)}_\alpha(f(\ell))))$. By Hypothesis and (\ref{eq: claim}), we have two main cases:

\noindent Case (a). $S^{(g,f)}_\alpha(f(\ell)))\in \widehat{Q}_{Ar}$ for some $1\leq r \leq n$. Then, $S^{(g,f)}_\alpha(f(\ell)))= \mathcal{A}r_q$ for some $q \in Q_{Ar}$. In that case, $\lder{\sigma}(\mathcal{A}r_q)\in \widehat{P}_{Ar}$. By (\ref{eq: general image of fAk}), $f(\lder{\sigma}(\mathcal{A}r_q))\in \bigcup_{k=1}^n \widehat{Q}_{Ak}$. Thus, $S^{(g,f)}_{\alpha\sigma}(f(\ell))\in \bigcup_{k=1}^n \widehat{Q}_{Ak}$, and the property holds.\\
\noindent Case (b). $S^{(g,f)}_\alpha(f(\ell)))\in Q^{(g_{Aj},f_{Aj})}_{f_{Aj}(\ell)}$. By the structure of this set of states given above, we have two subcases:
\begin{quote}
Case (b1). $S^{(g,f)}_\alpha(f(\ell)))\in \widehat{Q}_{Aj}$. The proof is the same as in Case (a).\\
Case (b2). $S^{(g,f)}_\alpha(f(\ell)))= \lder{\beta}(\ell)$ for some $\beta\in \Sast$. Thus, $f(\lder{\sigma}(\lder{\beta}(\ell)))= f(\lder{\beta\sigma}(\ell))$.
Then, by (\ref{eq: general image of fAk}), if $\lder{\beta\sigma}(\ell)\notin \bigcup_{k=1}^n \widehat{P}_{Ak}$, then $f(\lder{\beta\sigma}(\ell))= \lder{\beta\sigma}(\ell)$. As $\lder{\beta\sigma}(\ell)\notin \widehat{P}_{Aj}$, then $\lder{\beta\sigma}(\ell)=f_{Aj}(\lder{\beta\sigma}(\ell))= f_{Aj}(\lder{\sigma}(\lder{\beta}(\ell))) \in Q^{(g_{Aj},f_{Aj})}_{f_{Aj}(\ell)}$. Then, $S^{(g,f)}_{\alpha\sigma}(f(\ell))\in Q^{(g_{Aj},f_{Aj})}_{f_{Aj}(\ell)}$ and the property holds. Finally, by (\ref{eq: general image of fAk}), if $\lder{\beta\sigma}(\ell)\in \bigcup_{k=1}^n \widehat{P}_{Ak}$, then $f(\lder{\beta\sigma}(\ell))\in \bigcup_{k=1}^n \widehat{Q}_{Ak}$. In this case, $S^{(g,f)}_{\alpha\sigma}(f(\ell))\in \bigcup_{k=1}^n \widehat{Q}_{Ak}$ and the property holds again.
\end{quote}
Therefore, $\{ S^{(g,f)}_\alpha(f(\ell))~| \alpha\in \Sast\}\subseteq (\bigcup_{k=1}^{n} \widehat{Q}_{Ak}) \cup Q^{(g_{Aj},f_{Aj})}_{f_{Aj}(\ell)}$. This implies that it is finite. By Lemma \ref{theo: recognizability}, $f(\ell)\in L$ is recognized by the $M$-DFA $N^{(g,f)}(f(\ell),1)$. In conclusion, $\ell \in \mbox{RcgCap}^{(g_{Aj},f_{Aj})}_L$ implies $\ell \in \mbox{RcgCap}^{(g,f)}_L$; and, as $\ell$ and $Aj$ has been selected in an arbitrary way, then (\ref{eq: lemma composition nat fact}) holds. \hfill \qed

\section{Conclusions}\label{sec: conclusions}

The Myhill-Nerode theory studies formal languages and deterministic automata through right congruences and congruences on a free monoid. Let $(M,\cdot, 1)$ be an arbitrary monoid. In this paper, we provide a general Myhill-Nerode theorem for $M$-languages, i.e., functions of the form $\ell:\Sast \rightarrow M$. $M$-languages are studied from the aspect of their recognition by deterministic finite automata whose components take values on $M$ ($M$-DFAs). Unlike other previous papers in the literature that deal with the problem of recognizability of languages on different algebraic structures, we do not assume any additional property on the monoid. We characterize an $M$-language $\ell$ by a right congruence on $\Sast$ that is defined through the language $\ell$ and a factorization $(f,g)$ on the set of all $M$-languages, denoted $\equiv^{(g,f)}_\ell$. As $(g,f)$ is a factorization then $\ell= g(\ell)\cdot f(\ell)$ where $g(\ell)\in M$ and $f(\ell)$ is an $M$-language. Then, $\ell$ is characterized equivalently by both $\equiv^{(g,f)}_\ell$ or $\equiv^{(g,f)}_{f(\ell)}$ congruences. The main result of the paper (Theorem \ref{theo: main result}) states the equivalence of the conditions:
\begin{quote}
(i) $\ell$ is a recognizable $M$-language;\\
(ii) there exists a factorization on $L$, $(g,f)$, such that the right congruence $\equiv_{f(\ell)}^{(g,f)}$ has finite index.
\end{quote}
The proof of the implication (ii)$\Rightarrow$(i) requires the explicit construction of an $M$-DFA that recognizes $f(\ell)$. The properties of this automata based on the congruence $\equiv_{f(\ell)}^{(g,f)}$ state that it is accessible and satisfies a weak necessary condition of minimality (Property \ref{prop: prop of Nerode auto}). The proof of the implication (i)$\Rightarrow$(ii) requires that any $M$-DFA induces a factorization on the set of all $M$-languages. That factorization is called \emph{natural factorization} induced by an $M$-DFA. The construction and composition of this kind of factorizations is also studied in the paper. The provided formalism to study factorizations and their composition establishes that the composition of natural factorizations is also a factorization that preserves the recognition capability of each individual natural factorization (Lemma \ref{lemma: composition of natural fact}). In the particular case of the existence of a maximal factorization on the set of all $M$-languages, we obtain that the automata based on a maximal factorization is minimal (Property \ref{prop: mimimal by maxfact}) and that the recognition capability is maximal (Lemma \ref{lem: recgacp by maxfact}).



\begin{thebibliography}{99}
\bibitem{Belohklavek2002} R. B\v{e}lohl\'avek. Determinism and fuzzy automata. Information Sciences 143 (2002) 205--209.
\bibitem{Bozapalidis2008} S. Bozapalidis, O. Louscou-Bozapalidou. On the recognizability of fuzzy languages II. Fuzzy Sets and Systems 159 (2008) 107--113.
\bibitem{Brzozowski1962} J. A. Brzozowski. Derivatives of Regular Expressions. Journal of ACM. 11:4 (1964) 481-–494.
\bibitem{Ciric2010}  M. \'Ciri\'c, M. Droste, J. Ignjatovi\'c, H. Vogler. Determinization of weighted finite automata over strong bimonoids. Information Sciences 180 (2010) 3497--3520.
\bibitem{Droste2009} M. Droste, W. Kuich, and H. Vogler, editors. Handbook of Weighted Automata. Springer-Verlag, Berlin, 2009.
\bibitem{Droste2013} M. Droste,W. Kuich, Weighted finite automata over hemirings, Theoretical Computer Science 485 (2013) 38--48.
\bibitem{Droste2010} M. Droste, T. St\"uber, H. Vogler. Weighted finite automata over strong bimonoids. Information Sciences 180 (2010) 156--166.
\bibitem{Eilenberg1974} S. Eilenberg. Automata, Languages and Machines. Academic Press. New York and London 1974.
\bibitem{Eisner2003} J. Eisner. Simpler and more general minimization for weighted finite-state automata. In Proceedings of HLT-NAACL2003 conference, pp. 64--71, 2003.
\bibitem{Gerdjikov2017} S. Gerdjikov, S. Mihov. Myhill-Nerode Relation for Sequentiable Structures. ArXiv e-prints. https://arxiv.org/abs/1706.02910, (2017).
\bibitem{Gerdjikov2018} S. Gerdjikov. A general class of monoids supporting canonisation and minimisation of (sub)sequential transducers. Klein, S.T., Mart\'in-Vide, C., Shapira, D. (eds.) Language and Automata Theory and Application, 12th International Conference LATA2018 (2018).
\bibitem{Gerdjikov2018b} S. Gerdjikov. Characterisation of (sub)sequential rational function over a general class monoids. CoRR, abs/1801.10063, (2018).
\bibitem{Gerdjikov2019} S. Gerdjikov, J. R. Gonz\'alez de Mendivil. Conditions for the existence of maximal factorizations. Fuzzy Sets and Systems. (on line) https://doi.org/10.1016/j.fss.2019.07.006 (2019).
\bibitem{Mendivil2014} J. R. Gonzalez de Mendivil, J. R. Garitagoitia. Fuzzy languages of infinite range: pumping lemmas and determinization procedure. Fuzzy Sets and Systems 249 (2014) 1--26.
\bibitem{Mendivil2014b} J. R. Gonzalez de Mendivil, J. R. Garitagoitia. Determinization of fuzzy automata via factorization of fuzzy states. Information Sciences 283 (2014) 165--179.
\bibitem{mendivil2016} J. R. Gonz\'alez de Mendivil. A generalization of Myhill-Nerode theorem for fuzzy languages. Fuzzy Sets and Systems 301 (2016) 103--115.
\bibitem{mendivil2018} J. R. Gonz\'alez de Mendivil. Conditions for Minimal Fuzzy Deterministic Finite Automata via Brzozowski's Procedure. IEEE Transactions on Fuzzy Systems. 26 (4) (2018) 2409--2420.
\bibitem{mendivil2018b} J. R. Gonz\'alez de Mendivil, F. Fariña. Canonization of max-min fuzzy automata. Fuzzy Sets and Systems. 376 (2019) 152--168.
\bibitem{fitxi2019} J. R. Gonz\'alez de Mendivil, F. Fari\~na. Recognizability of languages with values on a monoid. Report number: M02-2019-gsd. Universidad P\'ublica de Navarra. 2019.
\bibitem{Hopcroft2007} J. E. Hopcroft, R. Motwani, J. D. Ullman. \emph{Introduction to Automata Theory}. 3rd Edition. Addison-Wesley, 2007.
\bibitem{Ignjatovic2008} J. Ignjatovi\'c, M. \'Ciri\'c, S. Bogdanovi\'c. Determinization of fuzzy automata with membership values in complete residuated lattices. Information Sciences 178 (2008) 164--180.
\bibitem{Jelena2010} J. Ignjatovi\'c, M. \'Ciri\'c, S. Bogdanovi\'c, T. Petkovi\'c. Myhill-Nerode type theory for fuzzy languages and automata. Fuzzy Sets and Systems 161 (2010) 1288--1324.
\bibitem{Kirsten2005} D. Kirsten, L. M\"aurer. On the determinization of weighted automata. Journal of Automata, Languages and Combinatorics 10 (2005) 287--312.
\bibitem{Klement2000} E.P. Klement, R. Mesiar and E. Pap. Triangular Norms, Kluwer Academic Publishers, Dordrecht, 2000.
\bibitem{Li2005} Y. M. Li, W. Pedrycz, Fuzzy finite automata and fuzzy regular expressions with membership values in lattice ordered monoids. Fuzzy Sets and Systems 156 (2005) 68--92.
\bibitem{Li2007} Y. M. Li, W. Pedrycz. Minimization of lattice finite automata and its application to the decomposition of lattice languages. Fuzzy Sets and Systems 158 (2007) 1423--1436.
\bibitem{Li2011} Y. Li. Finite automata theory with membership values in lattices. Information Sciences 181 (2011) 1003--1017.

\bibitem{Mohri1997} M. Mohri. Finite-state transducers in language and speech processing. Computing Linguistics 23(2) (1997) 269--311.
\bibitem{Mohri2000} M. Mohri. Minimization algorithms for sequential transducers. Theoretical Computer Science 234 (2000) 177--201.
\bibitem{Mordeson2002} J. Mordeson, D. Malik. \emph{Fuzzy Automata and Languages: Theory and Applications}. Chapman \& Hall, CRC Press, London, Boca Raton, FL., 2002.
\bibitem{Myhill1957} J. Myhill. Finite automata and the representation of events. WADD TR-57-624, Wright Patterson AFB, Ohio, pp. 112--137, 1957.
\bibitem{Nerode1958} A. Nerode. Linear automata transformation. Proceedings of AMS 9, pp. 541--544, 1958.

\bibitem{Petkovic2005} T. Petkovi\'c. Varietes of fuzzy languages. Proc. 1st Internat. Conf. on Algebraic Informatics, Aristotle University of Thessaloniki, Thessaloniki, pp. 197--205, 2005.
\bibitem{Qiu2001} D.W. Qiu. Automata theory based on completed residuated lattice-valued logic (I). Science in China, Ser. F, 44 (6) (2001) 419--429.
\bibitem{Qiu2002} D.W. Qiu. Automata theory based on completed residuated lattice-valued logic (II). Science in China, Ser. F, 45 (6) (2002) 442--452.
\bibitem{Rahonis2009} G. Rahonis. Fuzzy languages. Handbook of Weighted Automata. M. Droste, W. Kuich, H. Vogler (Eds.), Springer-Verlag, Berlin, 2009.
\bibitem{Sakarovitch2009} J. Sakarovitch. Elements of Automata Theory. Cambridge University Press. 2009.
\bibitem{Shen1996} J. Shen. Fuzzy language on free monoid. Information Sciences 88 (1996) 149--168.
\bibitem{Souza2000} R. Souza. Properties of some classes of rational relations (short version in English). Master’s thesis, University of Sao Paulo, 2004.
\bibitem{Stanimirovic2017} S. Stanimirovi\'c et. al. Determinization of fuzzy automata by factorizations of fuzzy states and right invariant fuzzy quasi-orders. Information Sciences 469 (2018) 79--100.
\bibitem{Stanimirovic2017b} S. Stanimirovi\'c et. al. A double reverse canonization method for fuzzy automata. Internal communication 2017.
\bibitem{Zadeh1965} L. A. Zadeh. Fuzzy Sets. Information and Control 8 (3) (1965) 338--353.


\end{thebibliography}
\end{document}